# Phonon modes and topological phonon properties in $(GaN)_x/(AlN)_x$ and $(AlGaN)_x /(GaN)_x$ superlattices

Dao-Sheng Tang [1,2,*], Li-Min Zhang [3]

[1] School of Rail Transportation, Soochow University, Suzhou 215006, China;

[2] School of Physical Science and Technology, Soochow University, Suzhou 215006, China;

[3] The Second Affiliated Hospital of Soochow University, Suzhou 215006, China

*Corresponding author, Email: dstang@suda.edu.cn

(July 11, 2023)

**Abstract:** To effectively regulate thermal transport for the near-junction thermal management of GaN electronics, it is imperative to gain an understanding of the phonon characteristics of GaN nanostructures, particularly the topological phonon properties connected to low-dissipation surface phonon states. In this work, a comprehensive study on phonon modes and topological phonon properties is performed from first principles in $(GaN)_x/(AlN)_x$ and $(AlGaN)_x/(GaN)_x$ ($x$=1,2,3) superlattices. Phonon modes, including the dispersion relation, density of states, and participation ratio, were calculated for six GaN superlattices. The participation ratio results did not reveal the localized phonon mode. In topological phonon analyses, it is found that Weyl phonons with a Chern number of 1(-1) are present in all six GaN superlattices, consisting of trivial (GaN) and nontrivial (AlN and AlGaN) combinations. These phonons are located on either side of the $k_z = 0$ plane symmetrically in the Brillouin zone. With the increase in the number of phonon branches in superlattices, the number of Weyl phonon points also increases from dozens to hundreds. One Weyl phonon with significant and clean surface states is selected and analyzed for each GaN superlattice. Among them, the Weyl phonon in $(GaN)_2/(AlN)_2$ superlattice mainly results from the lattice vibrations of Al and Ga atoms, while the Weyl phonons in other superlattices mainly result from the lattice vibrations of N atoms. The Weyl phonons at opposite $k_z$ planes form pairs in $(GaN)_2/(AlN)_2$, AlGaN/GaN, and $(AlGaN)_2/(GaN)_2$. Effects of strain including biaxial and uniaxial strain on Weyl phonons in GaN/AlN and AlGaN/GaN superlattices are investigated. Results indicate that Weyl phonons persist in large strain states, however, no monoclinic trend is observed due to the accidental degeneracy of these superlattices. The investigation in this work is promising to provide a deeper understanding of phonon properties and the topological effects of phonons in GaN nanostructures.

# I. Introduction

Gallium Nitride (GaN), along with a series of related nitride semiconductors, has become the most important power semiconductor due to its excellent electronic properties, including a wide bandgap, high electron mobility, high breakdown electric fields, and high saturated current. [1,2]. Accompanied by the increase in power performance including higher power density and working frequency, the heat dissipation issue has become a bottleneck problem for the development of better electronics [3-8]. Recent investigations show that thermal management near the junction is the most important factor in dissipating electronic heat. Specifically, there are three important characteristics in the near-junction heat conduction process, including the coupling between heat generation and heat conduction, phonon ballistic-diffusive transport, and thermal spreading [5-7]. Hence, understanding the phonon properties, especially those of nitride nanostructures, is essential.

Inspired by research on topological physics in electron and photon systems, research on topological phonons, *i.e.*, the geometric properties of phonon eigenmodes, is increasing rapidly and shows the possibility of realizing non-dissipative surface phonon transport [9-11]. This also promises to provide a solution for the near-junction heat dissipation issue, such as increasing the heat-spreading efficiency at the interface. In the past several years, researchers have found abundant topological phonon states in typical semiconductor lattice systems [11-15], such as GaN, Si, graphene, and SiO2. Nodal chain phonons and Weyl phonon topological phase transitions have been reported in wurtzite GaN systems [16,17]. Since semiconductor materials with nontrivial topological phonon characteristics will be used as nanostructures or parts of composite structures in both electronics and thermal functional materials such as nanofilms, nanowires, heterostructures, and superlattices, i.e., integrated with other materials hosting trivial or nontrivial topological phonon properties, it is of great interest in both theoretical and practical research to explore how the topological phonon properties change in these cases. Besides, superlattices and heterostructures are common nanostructures used for thermal functional materials and electronics. For example, the AlGaN/GaN heterostructure is the core structure in GaN high electron mobility transistors, and the GaN/AlN heterostructure (superlattice) is commonly present in the area where GaN and the substrate are bonded. However, composite nanostructures beyond single crystals are still under development for topological phonons.

The superlattice structure is not new to the scientific community and has been studied for many years due to its extraordinary electronic and optical properties [18-23]. It is also an ideal platform to investigate novel physical effects. Due to limitations in the development of *ab initio* methods and computing resources, force constant models and macroscopic approximation models were often adopted to study phonon properties in GaN-based superlattice structures in early research [24-32]. Due to the expansion of the unit cell, more phonon modes are generated, including interface modes, quasi-confined modes, and extended modes. First-principles calculations provide more detailed phonon information at the atomic scale and accurate optical properties of GaN superlattices [33-35]. In the thermal transport research field, superlattice structures are supposed to decrease the phonon mean free path with the interface phonon scatterings and tune the phonon dispersions owing to the phonon coherence effect [36-38]. These represent the two aspects of understanding the phonon transport in superlattice structures, *i.e.*, the particle and wave nature of phonons. The most recent investigation results wherein the GaN system hosts nontrivial topological phonon properties [16,17] as mentioned above, open more channels to understand the phonon properties in GaN superlattice structures. Exploring the topological phonons in composite GaN nanostructures will particularly enrich our current understanding.

In this work, two kinds of superlattice structures including $(GaN)_x/(AlN)_x$ and $(AlGaN)_x/(GaN)_x$ ($x$=1,2,3) superlattices are investigated from first principles focusing on their phonon modes and topological phonon properties. Phonon dispersion relations, along with the densities of states and participation ratios, are provided for all six GaN superlattice structures. In particular, dozens of single-type Weyl phonons are predicted in these superlattice structures, as confirmed by the results of the Berry curvature and the Chern number. By selecting one Weyl phonon with significant and clean surface states for each superlattice structure, the Weyl phonons are visualized with eigenvector information and analyzed in detail with the surface states and the local density of states. In addition, the strain effects on the six superlattices are investigated, analyzing the distributions of Weyl phonons in the respective irreducible Brillouin zones.

## II. Theory and methods

### A. Lattice dynamics and phonon modes

In this work, $(GaN)_x/(AlN)_x$ and $(AlGaN)_x/(GaN)_x$ superlattices, as well as GaN, AlN,

and wurtzite-type AlGaN crystals, are investigated using first principles calculation methods and lattice dynamics theory. For a three-dimensional lattice with more than one atom (*n*) in each unitcell, the equations of motion for the atoms can be described as

$$m(jl)\ddot{u}_\alpha(jl) = -\sum_{lj\alpha} \Phi_{\alpha\alpha'}(jl,j'l')u_{\alpha'}(j'l') \quad (1)$$

where $\Phi_{\alpha\alpha'}$ is the second-order interatomic force constants (2nd IFCs) defined as

$$\Phi_{\alpha\beta}(jl,j'l') = \frac{\partial^2 V}{\partial r_\alpha(jl)\partial r_\beta(j'l')} = -\frac{F_\beta(j'l')}{\partial r_\alpha(jl)} \quad (2)$$

In this formula, *V* is the potential energy of the phonon system, a function of atomic positions, *F* is the force on the atom which will be calculated directly from first principles. *α* and *β* are the Cartesian indices, *j* and *j'* are the indices of atoms in a unitcell, and *l* and *l'* denote the indexes of unitcells. With a lattice wave solution and a Fourier transformation from real space to momentum space, the lattice dynamics can be determined by the following equation

$$\sum_{j'\beta} D_{\alpha\beta}(jj',\mathbf{q})e_\beta(j',\mathbf{q}\upsilon) = [\omega(\mathbf{q}\upsilon)]^2 e_\alpha(j,\mathbf{q}\upsilon) \quad (3)$$

where **q** and *υ* are phonon wave vector and branch, and *D* is the dynamic matrix defined as [39]

$$D_{\alpha\beta}(jj',\mathbf{q}) = \sum_{l'} \frac{\Phi_{\alpha\beta}(0j,l'j')}{\sqrt{m_j m_{j'}}} e^{i\mathbf{q}\cdot[\mathbf{r}(l'j')-\mathbf{r}(0j)]} \quad (4)$$

And its general form is

$$\mathbf{D}(\mathbf{q})\mathbf{e}_{\mathbf{q}\upsilon} = \omega_{\mathbf{q}j}^2 \mathbf{e}_{\mathbf{q}\upsilon} \quad (5)$$

where **D** is a 3*n*×3*n* matrix, and **e** is the phonon eigenvector. For the long-range interaction of the macroscopic electric field induced by the polarization of collective ionic motions near the Γ point, which is commonly present in polar materials, a non-analytical term is required to correct the dynamics matrix. In this work, the Born effective charges *Z*\* and the dielectric tensor *ε* were calculated using density functional perturbation theory to modify the force constants in polar crystals [40-42].

The 2nd IFCs are calculated based on the finite displacement method from first principles as implemented in the open-source package Phonopy [43]. For single crystal lattices (GaN, AlN, AlGaN), 4×4×3 supercells containing 192 atoms were used in force calculations. The *k*-point grids for Brillouin zone integration were changed to 2×2×2 to be commensurate with the supercell dimensions. For superlattice structures $(GaN)_x/(AlN)_x$ and $(AlGaN)_x/(GaN)_x$, 4×4×2 supercells containing 256 atoms and 512 atoms were adopted for cases with $x$ equal to 1 and 2. 3×3×2 supercells containing 432 atoms were used for six-layer superlattices, *i.e.*, cases where $x$ equals 3. For the supercell calculations of superlattice structures, the *k*-point grid for Brillouin zone integration was set to 2×2×1. All first-principles calculations in this work were performed using the open-source software Quantum ESPRESSO [44]. The Perdew-Burke-Ernzerhof (PBE) of the generalized gradient approximation [45] was employed for the exchange-correlation functional. The projector augmented wave (PAW) [46] pseudopotentials were employed to treat the Ga, N, and Al atoms. A kinetic energy cutoff value of 60 Ry and 8×8×6 Monkhorst-Pack *k*-point grids [47] was applied to the Brillouin zone integration for structural optimization, with a tight convergence criterion of $10^{-8}$ Ry/Bohr for the Hellman-Feynman forces acting on each atom, and $10^{-8}$ Ry for the total energy. Convergence tests for *k*-mesh grids and energy cutoffs are performed for each of the six superlattice structures, and can be found in the Supplementary Material.

Considering that phonon localization may be present in superlattice structures, the mode participation ratio (PR) is adopted to determine the degree of localization of each phonon mode. PR is defined as [48-50]

$$P_\lambda = \frac{1}{N \sum_i (\sum_\alpha e^*_{i\alpha,\lambda} e_{i\alpha,\lambda})^2} \tag{6}$$

or

$$P_\lambda = \frac{(\sum_i (\sum_\alpha e^*_{i\alpha,\lambda} e_{i\alpha,\lambda}))^2}{N \sum_i (\sum_\alpha e^*_{i\alpha,\lambda} e_{i\alpha,\lambda})^2} \tag{7}$$

where $N$ is the total number of atoms in the research system (a unitcell in this work), $e_{i\alpha,\lambda}$ is the $\alpha$th eigenvector component of eigenmode $\lambda$ for the *i*th atom, and the superscript * denotes the complex conjugate.

**B. Topological phonon theory**

The topological phonon property, similar to its correspondence in the electronic system, is the property of phonon eigenvectors. By solving the phonon eigenvalue equation, both the eigenvalues and eigenvectors can be obtained. And the Berry connection can be then defined as [9,10]

$$\mathbf{A}_{n,\mathbf{k}} = -i\langle \mathbf{u}_{n,\mathbf{k}}|\nabla_\mathbf{k}|\mathbf{u}_{n,\mathbf{k}}\rangle \tag{8}$$

The Berry curvature is defined as the curl of the Berry connection.

$$\mathbf{\Omega}_{n,\mathbf{k}} = \nabla_\mathbf{k} \times \mathbf{A}_{n,\mathbf{k}} \tag{9}$$

The Berry phase is the integral of the Berry connection along a specific path, usually a circular path,

$$\gamma_n = \int_L \mathbf{A}_{n,\mathbf{k}} \cdot d\mathbf{k}. \tag{10}$$

The topological quantity for Weyl phonon states is the Chern number, which is defined on a closed surface containing a Weyl point in reciprocal space.

$$C = \frac{1}{2\pi} \oint \Omega_{n,\mathbf{k}} d^2\mathbf{k}. \tag{11}$$

The characteristics of a nontrivial topological phonon state can be illustrated by the Berry curvature, Berry phase, and Chern number. Moreover, the evolution of the Wannier charge center (WCC) is also a significant indicator [11,14]. The Wannier charge centers ($\varphi$) are defined on the series of Wilson loops (parameterized by $\theta$) over a sphere enclosing a Weyl phonon point. For a single-type Weyl point, $\varphi$ changes from zero to $2\pi$ as $\varphi$ varies from 0 to $\pi$ [51]. The physical quantities introduced above are all bulk quantities directly derived from the bulk phonon eigenvectors. Since the significant topological feature is the bulk-boundary correspondence, the surface phonon states are the most direct evidence of phonon topology. Generally, surface phonon states can be extracted from both calculations on a slab system and calculations using the Green's function method, as implemented in the open-source software WannierTools [51].

### III. Results and discussion

### A. Lattice structures of superlattices

The relaxed structures are shown in Figures 1-3. GaN and AlN are both of a wurtzite structure with the space group *P*6$_3$*mc* (No. 186). The lattice symmetry decreases due to the mixing of Al and Ga atoms in the wurtzite-type AlGaN, with the space group *P*3*m*1 (No. 156). With the same structure and similar lattice parameters, the matching degree between GaN and AlN (AlGaN$_2$) is quite high; only the lattice parameters change in the superlattice structures, while the wurtzite type structure, as well as the space group *P*3*m*1, are still conserved. Detailed lattice parameter information is present in Table 1. According to the first-principles data, GaN has the largest lattice parameter in the in-plane direction (perpendicular to the polar axis), while AlN has the smallest one. Lattice parameters in superlattice structures are smaller than those in GaN and larger than those in AlN (AlGaN). Therefore, GaN cells in the superlattice structure are stressed with compressive strain while AlN (AlGaN) cells are with tensile strain. In general, the dependence of lattice parameters on the thickness of the unit cell is not significant. For $(GaN)_x/(AlN)_x$ superlattice, the in-plane lattice parameter $a$ increases with the increase of the unit cell thickness of the superlattice, while the normalized out-of-plane lattice parameter $c$ (the actual $c$ divided by the number of layers of the original wurtzite unit cell) decreases. The relative changes are small, less than 0.06%. The actual Ga-N length along the $c$-axis significantly increases in the superlattice, while the Al-N length decreases. The $(AlGaN)_x/(GaN)_x$ superlattice shows similar changes quantitatively, but with much smaller values.

Though the basic lattice structure is conserved and the space group does not change with the increase of the superlattice thickness, changes in lattice parameters and atomic lengths are promising to renormalize the force constants and induce variations in the phonon dispersions, *e.g.*, an increase in the number of flat bands and crossing points. This may provide more candidate phonon points with nontrivial topological properties, as discussed in the later section III. C.

**B. Phonon dispersion and modal localization**

Superlattice structures investigated in this work consist of two, four, and six original unit cells, respectively, corresponding to eight, sixteen, and twenty-four atoms in each new unit cell, which then generate more phonon branches. These modes may include both non-local bulk modes and local interfacial modes. Phonon dispersions and densities of state (DOS) of $(GaN)_x/(AlN)_x$ and $(AlGaN)_x/(GaN)_x$, as well as of GaN, AlN, and AlGaN single crystals, are shown in Figures 4-6. As illustrated in Figure 4, GaN and its related single-crystal nitride semiconductors (AlN and AlGaN single crystals) show similar phonon dispersions, with LO-TO splitting and obvious gaps between the high and low phonon branches [52,53]. Due to the difference in atomic mass, N atoms mainly contribute to the lattice vibration at high frequencies; for example, the projected DOS of the N atom is nearly zero at the low-frequency part in GaN, while it contributes to almost 100% at the high frequencies above the gap. The atomic mass of the Al atom is in between that of the Ga and N atoms, therefore, the phonon frequencies in AlN and AlGaN are higher than those in GaN, and the projected DOS in AlN shows a more uniform distribution. In AlGaN, Ga atoms mainly contribute to the lattice vibration around a frequency of 5 THz, Al atoms contribute to the lattice vibration around a larger frequency range, around 12 THz, while N atoms dominate in the highest range, around 20 THz. The different projected DOS of different atoms also account for the participation ratios in the single crystals of GaN, AlN, and AlGaN. The participation ratio indicates the localization degree of the phonon modes, and is promised to be or close to 1 in single crystals since all phonon modes are propagons instead of locons or diffusons [54,55]. For single crystals with only one kind of element atom, such as silicon, the PR is nearly equal to one for all phonon modes. Due to the spectral projected DOS properties in GaN, AlN, and AlGaN, the PRs are slightly less than one, as seen in Figure 7.

In a superlattice structure, phonon branches become much more numerous than those in single crystals, with more phonon bands crossing and degenerating with each other, forming flat phonon bands. As illustrated by Figures 5 and 6, with the increase of the original unit cell layers in superlattices, a significant concentration of frequencies appears in the phonon dispersion relations. Flat bands are also obviously present at frequencies of 18.5 THz for $(GaN)_x/(AlN)_x$ and 18 THz for $(AlGaN)_x/(GaN)_x$. Based on the PR criterion (PRs smaller than 0.1) for locons [54,55], there are no localized phonon states in all six GaN superlattice structures since PRs of all phonon modes are still larger than 0.1, with rarely a single data point in Figure 8. With the increase of the original unitcell layers in superlattices, PRs show a decrease, and localized phonon states are promising to be generated in GaN superlattices with larger unitcells.

**C. Topological phonon properties**

In this section, Weyl phonons are predicted for six GaN superlattice structures to analyze their topological phonon properties. With the aid of WannierTools [51], all phonon crossing points in the Brillouin zone of six GaN superlattice structures are searched. Further calculations of the Berry curvature, Chern number, and WCC evolution are carried out to determine the Weyl phonons, *i.e.*, phonon crossing points with nontrivial topological properties. It is found that, with the increase in the unit cell thickness of the superlattice, the number of Weyl phonons also increases, following the increase of phonon branches. Dozens of Weyl phonons are finally predicted in all six superlattice structures. Only taking into account the non-equivalent phonon wave vectors in the irreducible Brillouin zone, there are 19 Weyl phonon points in GaN/AlN superlattice, 97 in $(GaN)_2/(AlN)_2$, 311 in $(GaN)_3/(AlN)_3$, 21 in AlGaN/GaN, 125 in $(AlGaN)_2/(GaN)_2$, and 222 in $(AlGaN)_2/(GaN)_2$. These Weyl phonon points are shown in Table S1-S14 in the Supplementary Material containing information on phonon wave vectors, branches (bands), Chern number, and frequencies.

Previous studies on Weyl phonon points have shown that several kinds of Weyl phonons can present in three-dimensional lattice crystals without inversion symmetry including single type Weyl phonons with a Chern number of 1 (-1) [14,16], double Weyl phonons with a Chern number of 2 (-2) [56,57], Weyl complex (one double Weyl phonon connected by two single-type Weyl phonons) [15,56], charge-three and charge-four Weyl phonons with Chern numbers of 3 and 4 respectively [58]. For GaN crystal in particular, nodal chain phonons are present for all wurtzite structures with the space group $P6_3mc$ (No. 186) protected by the mirror symmetry [17]. And AlN and AlGaN have been reported to host Weyl phonons at non-high-symmetry points in the high-symmetry plane $k_z = 0$ while GaN does not [16]. The results in this work show that single-type Weyl phonons are formed in either high-symmetry paths or high-symmetry planes, but are located at both sides of the $k_z$=0 plane symmetrically in GaN superlattice structures.

Though dozens of Weyl phonons are found in these superlattice structures, their corresponding surface states may not be clean enough, or even hidden by other trivial surface states due to the large number of phonon branches in the superlattice structures. In order to demonstrate the Weyl phonon clearly, examples of superlattice structures that feature significant and clean surface states of the Weyl phonon will be discussed. The positions of selected Weyl phonons are illustrated in Figure 9. With the same lattice structures, all these six GaN superlattices host similar Brillouin zones, with only differences in magnitudes marked with blue lines in Figure 9, where Weyl phonons are marked with red dots. The length along the $k_z$ direction of a Brillouin zone decreases with the increase in the unitcell, resulting in smaller Brillouin zones. In each Brillouin zone of a superlattice structure, there are 12 equivalent Weyl phonons, with six of them in the $k_z = k_0$ plane (where $k_0$ is a constant and differs for different superlattices) and the remaining six in the $k_z = -k_0$ plane, corresponding to the combination of mirror symmetry and six-fold rotational symmetry. In general, Weyl phonons of a single type with double degeneracy are further classified into Type I and Type II Weyl points. To clearly illustrate the crossing types of Weyl phonons, three-dimensional phonon dispersions in the $k_x$-$k_y$ plane are plotted in Figure 10 for the selected Weyl phonons shown in Figure 9. It can be seen from Figures 10(a)-(f) that these six Weyl phonons, which belong to six different superlattice structures, are all Type-II Weyl phonons with linear and tilted phonon dispersions.

For all superlattice structures, detailed analyses are provided in the following paragraphs, with Figures 11-16. Weyl phonon positions, eigenvectors, and related topological phonon quantities in a superlattice GaN/AlN are shown in Figure 10. The Weyl phonons are located at the **q** point (0.399, 0.097, 0.032) and its equivalent points (Figure 9(a)), formed by the crossing of the phonon branches 22 and 23 at a frequency of 20.63 THz, where the phonon branches are sorted in ascending order. In Figure 11(a), the phonon dispersion relations along the path Γ-WP-K' are presented with a locally enlarged view showing the crossing point clearly. The label WP means Weyl phonon, and the label K' denotes the q point K but with the same $k_z$ as the Weyl phonon, *i.e.*, (1/3, 1/3, $k_z$). An enlarged view is shown in the upper right corner of Figure 11(a). It should be noted that the NAC term is not considered in the calculations of the Weyl phonons due to the current limit in the calculation algorithm. As a supplementary discussion, the phonon dispersion relations considering NAC are also plotted, accompanied by those without NAC, at the Weyl phonon points. The results show that the NAC does not fundamentally change the phonon band crossing, but mainly shifts the crossing positions slightly. The NAC mainly affects the IFC near the Γ point and is not supposed to significantly influence the phonon dispersion at large wavevectors. The detailed phonon eigenvector of a Weyl phonon is visualized in Figure 11(b) to understand its vibrational properties in real space. Basically, at non-Γ points, it is hard to distinguish the phonon polarizations simply as longitudinal or transverse. It can be deduced from the figure that the Weyl phonon is the coupled optical mode resulting from the atomic vibrations of GaN and AlN. Two N atoms (one in the GaN zone and the other in the AlN zone) vibrate with the largest magnitudes, followed by one Al atom with the third largest magnitude, while other atoms including one Al atom, two N atoms, and two Ga atoms show relatively small vibration magnitudes. The WCC evolution around a circle path enclosing the Weyl phonon and the Berry curvature distribution at $k_z = k_0$ ($k_0$=0.032 for this Weyl phonon) in the bulk plane, as shown in Figures 11(c) and (d), confirms the nontrivial topological phonon property of this phonon crossing point. The source and sink in the Berry curvature distribution denote the places where the Weyl phonons are located, marked with blue and red dots for Weyl phonons with Chern numbers 1 and -1 respectively. Figures 11(f) and (e) further illustrate the bulk-boundary correspondence of the Weyl phonon. The surface state at frequency 20.63 THz is projected into the surface Brillouin zone along the [0001] direction, and arc surface states are clearly shown in Figure 11(f) connecting the Weyl phonons with Chern numbers 1 and -1. The local DOS of surface phonon states along the path Γ-WP in Figure 11(e) also confirms the arc state where the surface phonon state of the Weyl phonon stops at the crossing point.

    With the increase of GaN and AlN atomic layers in the $(GaN)_x/(AlN)_x$ superlattice, the number of Weyl phonons increases from dozens to hundreds. However, the distribution of Weyl phonons is still similar. Two Weyl phonons in $(GaN)_2/(AlN)_2$ and $(GaN)_3/(AlN)_3$ are shown in Figures 9(b) and (c). For the superlattice $(GaN)_2/(AlN)_2$, the Weyl phonons are located at the **q** point (0.243, 0.141, 0.463) and its equivalent points, formed by the crossing of phonon branches 14 and 15 at a frequency of 8.56 THz. In Figure 12(a), the phonon dispersion relations along the path Γ-WP-K' are presented with a locally enlarged view showing the crossing point clearly. As seen in Figure 12(b), this Weyl phonon is also the result of the atomic vibrations in both the GaN and AlN zones. Different from that in the GaN/AlN structure, the Weyl phonon in $(GaN)_2/(AlN)_2$ evolves the vibrations of Al, Ga, and N atoms, with all three kinds of atoms showing relatively large vibration magnitudes, with the Al atoms vibrating with the largest magnitude. Figures 12(c)-(f) show the topological physical quantities of the Weyl phonon in the superlattice $(GaN)_2/(AlN)_2$. The blue dot in the Berry curvature distribution denotes the source position at the $k_z = -k_0$ plane, which is coupled with the Weyl point marked by the red dot at the $k_z = k_0$ plane. The arc surface states are projected in the surface Brillouin zone along the [0001] direction, with the local DOS plotted along the path connecting a couple of Weyl phonons in the surface Brillouin zone. It should be noted that the arc surface state at the surface Brillouin zone is the connection between two Weyl phonons with opposite $k_z$ components, different from the case in the former GaN/AlN superlattice structure. This phenomenon will also be found in AlGaN/GaN and $(AlGaN)_2/(GaN)_2$ superlattice structures, which will be discussed later. For the superlattice $(GaN)_3/(AlN)_3$, the Weyl phonons are located at the **q** point (0.043, 0.185, -0.027) and its equivalent points, formed by the crossing of the phonon branches 53 and 54 at a frequency of 18.65 THz. In Figure 13(a), the phonon dispersion relations along the path Γ-WP-M' are presented with a locally enlarged view showing the crossing point in detail. Eigenvector results in Figure 13(b) show that three N atoms

(one of them in the GaN zone and the other two in the AlN zone) vibrate with large magnitudes, while the magnitudes of other atoms are small, similar to the case in the GaN/AlN superlattice.

Basically, (AlGaN)$_x$/(GaN)$_x$ superlattice structures host the same lattice structure type as (GaN)$_x$/(AlN)$_x$ superlattices. The calculation results show that dozens to hundreds of Weyl phonons are also present in Brillouin zones, illustrated by Figures 14(d)-(f), with similar distributions. With the increase of the atomic layers in the superlattice, phonon branches increase, inducing many more surface states. In (AlGaN)$_x$/(GaN)$_x$ superlattice structures, there are also Weyl phonon points in the acoustic branches; however, the corresponding surface states cannot be distinguished directly due to the overlap of trivial and nontrivial modes, similar to the cases in (AlGaN)$_x$/(GaN)$_x$ superlattice structures. In Figure 14(a), phonon dispersion relations along the path Γ-WP-K' are shown where the Weyl phonons are located at the **q** point (0.291, 0.163, -0.313) and its equivalent points with a frequency of 17.51 THz. The phonon crossing points are formed by bands 14 and 15. This Weyl phonon is also an optical phonon, derived from the collective atomic vibration, where N atoms contribute the most to lattice vibrations and vibrate in opposite directions along the *z*-axis. In Figure 14(b), the N atom directly above the Al atoms in the AlGaN zone vibrates with the largest magnitude, followed by the N atoms in the GaN zone, and then the Ga atom and Al atom in the AlGaN zone. Figure 14(c) shows the typical characteristic of a single-type Weyl phonon. Similar to that in the superlattice (GaN)$_2$/(AlN)$_2$, Weyl phonons are formed into pairs with opposite *z* components. In Figure 14(d), a pair of Weyl phonons are marked with blue and red dots, indicating them as the source and sink in the Berry curvature distribution. The source labeled by the blue dot is not present at the $k_z = k_0$ plane but corresponds to the position at the $k_z = -k_0$ plane. The surface states with arc characteristics are also projected onto the surface Brillouin zones along the [0001] direction and their local DOS are plotted along the path connecting a pair of Weyl points, as shown in Figures 14(e) and (f). Slightly different from the Weyl phonons in other GaN superlattice structures, the Weyl phonons are connected by the surface states crossing the sides of the Brillouin zone. It is worth mentioning the phenomenon that a pair of Weyl phonons are connected by the surface states crossing the sides of the Brillouin zone of a wurtzite AlGaN single crystal [16]. For (AlGaN)$_2$/(GaN)$_2$, the Weyl phonons are located at the **q** point (0.217, 0.143, -0.495) and its equivalent points, very close to the top/bottom boundaries of the Brillouin zone, *i.e.*, $k_z = \pm 0.5$ plane. And they are formed by the crossing of bands 43 and 44 at a frequency of 20.14 THz. At the atomic scale, the vibrations of N atoms are dominant, showing much larger magnitudes than those of the Ga and Al atoms. Since the $k_z = k_0$ *and $k_z = -k_0$* planes are close to each other, the Weyl point as a source in the Berry curvature distribution at the $k_z=-k_0$ plane also shows the source characteristic in the $k_z=k_0$ plane, illustrated with a red dot in Figure 15(d). Corresponding arc surface states projected at the surface Brillouin zone along the [0001] direction and their local DOS can be seen in Figures 15(e) and (f). In (AlGaN)$_3$/(GaN)$_3$, the Weyl phonons are located at the **q** point (0.080, 0.231, 0.011) and its equivalent points, resulting from the phonon crossing of bands 50 and 51 at a frequency of 17.51 THz, as shown in Figure 16(a). The atomic vibrations at the AlGaN zone contribute the most to the Weyl phonon, while only one N atom at the GaN zone shows a significant vibration magnitude. At the AlGaN zone, three nonadjacent N atoms host vibrations with large magnitudes, and Ga and Al atoms do not. Topological physical properties are illustrated in Figures 16(c)-(f), confirming the nature of the single type of Weyl phonon and the presence of clear surface states.

As a short summary for this section, it is concluded that the selected six Weyl phonon points are all transverse optical phonons, though they are not strictly transverse optical ones with non-Γ wave vectors. Similar to the Weyl phonons reported in wurtzite semiconductors, such as AlN [16], CuI [59], and ZnSe [60], the Weyl phonons in GaN-related superlattice structures also result from the accidental degeneracy of phonon branches, and are not protected by symmetries. Accordingly, these Weyl phonons are also sensitive to perturbations, like lattice strain. Therefore, the strain effects on the distributions of Weyl phonons are illustrated in the following section.

**D. Strain effects on topological phonon properties**

In real GaN-based electronics, strain exists inevitably due to the mismatch between coherent layers and accurate control of electronic properties with strain engineering. In this section, Weyl phonons and their evolutions under four typical strain states are investigated for GaN/AlN and AlGaN/GaN superlattice structures. The strain is applied by directly changing the lattice constants in first-principles calculations. The four strain states are biaxial compressive, biaxial tensile, uniaxial compressive, and uniaxial tensile strain states. The relative change of 5% of lattice constants is adopted in each strain states, where the lattice constant *a* is changed in biaxial strain cases and lattice constant *c* is changed in uniaxial strain cases. The phonon dispersions for GaN/AlN and AlGaN/GaN superlattices under different strain states are illustrated in Figures 17 and 18. Changes in acoustic phonon branches are not significant while many phonon crossing points will be generated or disappear in optical branches with high phonon frequencies due to the lattice strain. The phonon DOSs of the superlattices under free state are present for reference in figures, which does not show large changes.

To figure out the evolution of topological phonon properties in GaN/AlN and AlGaN/GaN superlattices due to the lattice strain, non-equivalent Weyl phonon points in the irreducible Brillouin zones (IBZs) are illustrated for each lattice configuration. It is noted here that lattice symmetry is conserved in all strain states for both GaN/AlN and AlGaN/GaN superlattices. Seen from Figure 19, the Weyl phonons are mainly the phonon points with positive $k_z$ components and not located at the $k_z$=0 plane in the free state. No monotonous tendencies are observed with different strain states, while significant increase is shown for biaxial tensile strain state where the number of Weyl phonons in IBZ increases from 19 to 38 compared to that for free state. Besides, there is also no continuous change in the phonon crossing points; instead, new phonon crossing points are generated while the old ones disappear, since the strain applied in each case is relatively large. In Figure 20, the evolution of Weyl phonon points distribution in the AlGaN/GaN superlattice with strain states is illustrated, which is much different from that in the GaN/AlN superlattice. The four types of strain states all increase the Weyl phonon points in the IBZs, where the increases are especially significant for a biaxial tensile and a uniaxial compressive strain state.

The evolution of Weyl phonon points in these two superlattices with strain states shows the unstable property of Weyl phonons from the accidental degeneracy of phonon branches. However, Weyl phonon points remain existing in different cases, including the combination of a trivial and a non-trivial system, and superlattice structures under different strain states. Since the Weyl phonons in these wurtzite-type structures and relevant superlattice structures are due to the accidental degeneracy without protection of symmetries, the detailed and microscopic origin for the presence of Weyl phonons is expected to be revealed from the view of the interatomic force constants. This is an ongoing process and will be discussed in future work.

**IV. Conclusions and perspectives**

The phonon modal properties and topological effects of phonons in GaN superlattice structures are investigated from the first principles in this work. Six GaN superlattices including $(GaN)_x/(AlN)_x$ and $(AlGaN)_x/(GaN)_x$ with $x$ equal to 1, 2, and 3 are adopted as research objects to understand the phonon localization and Weyl phonons in GaN superlattice structures with different unitcells consisting of the trivial and nontrivial components. The calculations show that dozens of single-type Weyl phonons, with a Chern number of 1 (-1), are present in all six of the superlattices, including both acoustic and optical modes. They are not located in the high-symmetry plane, but at two sides of the $k_z = 0$ plane, symmetrically, in the Brillouin zone. For each superlattice structure, one Weyl phonon with significant and clean surface states is selected for analysis in the main text. All these six Weyl phonons, taken as examples, are shown to be the coupled modes in superlattices, rather than intrinsic GaN or AlN (AlGaN) phonon modes. The Weyl phonon in $(GaN)_2/(AlN)_2$ is mainly contributed by the vibrations of Al and Ga atoms, while the Weyl phonons in other superlattices are the result of the vibrations of N atoms in GaN and AlN (AlGaN) zones. And the Weyl phonons at opposite $k_z$ planes are formed into a pair in $(GaN)_2/(AlN)_2$, AlGaN/GaN, and $(AlGaN)_2/(GaN)_2$. Effects of strain, including biaxial and uniaxial strain, on Weyl phonons in GaN/AlN and AlGaN/GaN superlattices are investigated. Results indicate that Weyl phonons persist in large strain states; however, no monoclinic trend is observed due to the accidental degeneracy of these superlattices. In particular, a significant increase is shown for a biaxial tensile strain state, where the number of Weyl phonons in the IBZ increases from 19 to 38 compared to that for the free state in the GaN/AlN superlattice. All four types of strain states increase the Weyl phonon points in the IBZs, with the increases being especially significant for a biaxial tensile and a uniaxial compressive strain state in the AlGaN/GaN superlattice.

The Weyl phonons, as well as other topological phonons, with their unique transport properties, are promising to benefit research into thermal transport tuning, concerning the heat dissipation issues in electronic and functional material design. For example, couplings between electrons and phonons with non-trivial topological properties, the one-way propagation and phonon Hall effect [59] of Weyl phonons provide potential methods to tune the phonon thermal transport in the near-junction regions of electronics [5,6]. Besides, the transport of topological phonons in bulk materials may trigger different abnormal heat conduction mechanisms [61-63]. With comprehensive analyses of Weyl phonons in GaN-based superlattice structures, this work is promising to benefit the applications of the topological phonons in GaN nanostructures for electronic thermal management and thermal functional material design.

**Acknowledgments**

This work was financially supported by the National Natural Science Foundation of China (No. 52206105), the China Postdoctoral Science Foundation (No. 2021M702384), and Jiangsu Funding Program for Excellent Postdoctoral Talent (No. 2022ZB594).

**Author declarations**

The authors have no conflicts to disclose.

**Data availability**

The data that support the findings of this study are available from the corresponding author upon reasonable request.

Table 1 Lattice parameters of GaN related single crystals and superlattices

| | $a$ (Å) | $c$ (Å) | Normalized $c$ (Å) | $u$ (Ga-N) | $uc$ (Ga-N) | $u$ (Al-N) | $uc$ (Al-N) |
|---|---|---|---|---|---|---|---|
| GaN | 3.217 | 5.242 | - | 0.3767 | 1.9744 | - | |
| AlN | 3.132 | 5.021 | - | - | - | 0.3815 | 1.9156 |
| AlGaN | 3.1756 | 5.149 | - | 0.3858 | 1.9865 | 0.3708 | 1.9093 |
| GaN/AlN | 3.172 | 10.275 | 5.137 | 0.1929 | 1.9816 | 0.1858 | 1.9089 |
| (GaN)$_2$/(AlN)$_2$ | 3.174 | 20.542 | 5.136 | 0.0965 | 1.9826 | 0.0929 | 1.9086 |
| (GaN)$_3$/(AlN)$_3$ | 3.174 | 30.807 | 5.134 | 0.0644 | 1.9824 | 0.0620 | 1.9088 |
| AlGaN/GaN | 3.199 | 10.401 | 5.201 | 0.1905 | 1.9816 | 0.1834 | 1.9075 |
| (AlGaN)$_2$/(GaN)$_2$ | 3.199 | 20.802 | 5.201 | 0.0953 | 1.9816 | 0.0917 | 1.9073 |
| (AlGaN)$_3$/(GaN)$_3$ | 3.199 | 31.203 | 5.201 | 0.0635 | 1.9814 | 0.0611 | 1.9074 |

\* $u$ (Ga-N) and $u$ (Al-N) are both average values in superlattice structures.

**Figure captions**

**Figure 1** Lattice structures of GaN, AlN, and AlGaN. The lattice constants $a$, $c$, and internal parameter $u$ are illustrated in (a). The bright green dot denotes atom Ga, the blue-green dot denotes atom Al and the small gray dot atom N.

**Figure 2** Lattice structures of superlattices $(GaN)_x/(AlN)_x$. The bright green dot denotes atom Ga, the blue-green dot denotes atom Al and the small gray dot atom N.

**Figure 3** Lattice structures of superlattices $(AlGaN)_x/(GaN)_x$. The bright green dot denotes atom Ga, the blue-green dot denotes atom Al and the small gray dot atom N.

**Figure 4** Phonon dispersions and density of states of (a)(b) GaN, (c)(d) AlN, (e)(f) AlGaN

**Figure 5** Phonon dispersions and density of states of (a)(b) GaN/AlN, (c)(d) $(GaN)_2/(AlN)_2$, and (e)(f) $(GaN)_3/(AlN)_3$

**Figure 6** Phonon dispersions and density of states of (a)(b) AlGaN/GaN, (c)(d) $(AlGaN)_2/(GaN)_2$, and (e)(f) $(AlGaN)_3/(GaN)_3$

**Figure 7** Phonon modal participation ratio in GaN, AlN, AlGaN

**Figure 8** Phonon modal participation ratio in GaN superlattices: (a) $(GaN)_x/(AlN)_x$ and (b) $(AlGaN)_x/(GaN)_x$

**Figure 9** Brillouin zones of six GaN superlattice structures and the positions of Weyl phonons discussed in the main text. The black lines are used to indicate the Brillouin zone, blue lines for the $k_z=0$ plane, and high symmetry paths in this plane. The Weyl phonons are labeled with red dots.

**Figure 10** Three-dimensional phonon dispersions around the Weyl phonons in six superlattices marked in Figure 9.

**Figure 11** Weyl phonons at band 22 with **q** (0.399, 0.097, 0.032) at frequency 20.63 THz in GaN/AlN (a) Phonon dispersion (b) Phonon eigenvector (c) WCC evolution (d) Berry curvature (e) Surface arc state (f) Local destiny of states. The label K' in (a) denotes the **q** points K but with the same $k_z$ as the Weyl phonon, *i.e.*, (1/3, 1/3, $k_0$). The black line and red line in the enlarged view of (a) are the phonon dispersions at the Weyl phonon point without and with NAC. The vibration vectors in (b) are enlarged for a better view.

**Figure 12** Weyl phonons at band 14 with **q** (0.243, 0.141, 0.463) at frequency 8.56 THz in $(GaN)_2/(AlN)_2$ (a) Phonon dispersion (b) Phonon eigenvector (c) WCC evolution (d) Berry curvature (e) Surface arc state (f) Local destiny of states. The label K' in (a) denotes the **q** points K but with the same $k_z$ as the Weyl phonon, *i.e.*, (1/3, 1/3, $k_0$). The black line and red line in the enlarged view of (a) are the phonon dispersions at the Weyl phonon point without and with NAC. The blue dot in (d) illustrates the position of the Weyl phonon at the $k_z=-k_0$ plane.

**Figure 13** Weyl phonons at band 53 with **q** (0.043, 0.185, -0.027) at frequency 18.65 THz in $(GaN)_3/(AlN)_3$ (a) Phonon dispersion (b) Phonon eigenvector (c) WCC evolution (d) Berry curvature (e) Surface arc state (f) Local destiny of states. The label M' in (a) denotes the **q** points M but with the same $k_z$ as the Weyl phonon, *i.e.*, (1/3, 1/3, $k_0$). The black line and red line in the enlarged view of (a) are the phonon dispersions at the Weyl phonon point without and with NAC.

**Figure 14** Weyl phonons at band 15 with **q** (0.291, 0.163, -0.313) at frequency 17.51 THz in AlGaN/GaN (a) Phonon dispersion (b) Phonon eigenvector (c) WCC evolution (d) Berry curvature (e) Surface arc state (f) Local destiny of states. The label K' in (a) denotes the **q** points K but with the same $k_z$ as the Weyl phonon, *i.e.*, (1/3, 1/3, $k_0$). The black line and red line in the enlarged view of (a) are the phonon dispersions at the Weyl phonon point without and with NAC. The blue dot in (d) illustrates the position of the Weyl phonon at the $k_z=-k_0$ plane.

**Figure 15** Weyl phonons at band 43 with **q** (0.217, 0.143, -0.495) at frequency 20.14 THz in $(AlGaN)_2/(GaN)_2$ (a) Phonon dispersion (b) Phonon eigenvector (c) WCC evolution (d) Berry curvature (e) Surface arc state (f) Local destiny of states. The label M' in (a) denotes the **q** points M but with the same $k_z$ as the Weyl phonon, *i.e.*, (1/3, 1/3, $k_0$). The black line and red line in the enlarged view of (a) are the phonon dispersions at the Weyl phonon point without and with NAC. The sink marked by the blue dot in (d) Berry curvature is the indication of the Weyl phonon at the $k=-k_0$ plane, which is also present in the Berry curvature at the $k=k_0$ plane due to the short distance between these two planes.

**Figure 16** Weyl phonons at band 50 with **q** (0.080, 0.231, 0.011) at frequency 17.51 THz in $(AlGaN)_3/(GaN)_3$ (a) Phonon dispersion (b) Phonon eigenvector (c) WCC evolution (d) Berry curvature (e) Surface arc state (f) Local destiny of states. The label M' in (a) denotes the **q** points M but with the same $k_z$ as the Weyl phonon, *i.e.*, (1/3, 1/3, $k_0$). The black line and red line in the enlarged view of (a) are the phonon dispersions at the Weyl phonon point without and with NAC.

**Figure 17** Phonon dispersions of GaN/AlN superlattice under different strain states (a) Compressive and tensile biaxial strain, (b) Compressive and tensile uniaxial strain along the polar axis.

**Figure 18** Phonon dispersions of AlGaN/GaN superlattice under different strain states (a) Compressive and tensile biaxial strain, (b) Compressive and tensile uniaxial strain along the polar axis.

**Figure 19** Weyl phonon points in irreducible Brillouin zone (marked by the blue lines) for GaN/AlN superlattice under different strain states (a) free, (b) biaxial compressive strain, (c) biaxial tensile stain, (d) uniaxial compressive strain, (e) uniaxial tensile strain.

**Figure 20** Weyl phonon points in irreducible Brillouin zone (marked by the blue lines) for AlGaN/GaN superlattice under different strain states (a) free, (b) biaxial compressive strain, (c) biaxial tensile stain, (d) uniaxial compressive strain, (e) uniaxial tensile strain.

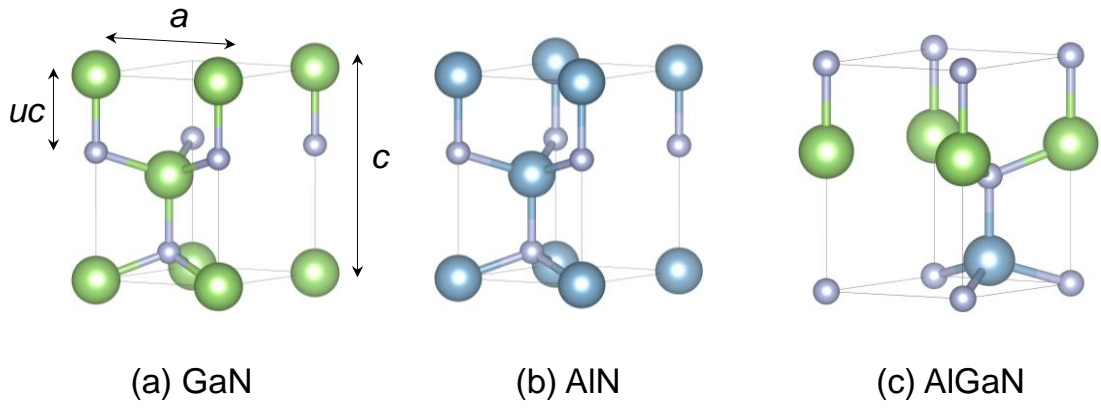

(a) GaN  (b) AlN  (c) AlGaN

Figure 1 Lattice structures of GaN, AlN, and AlGaN. The lattice constants *a*, *c*, and internal parameter *u* are illustrated in (a). The bright green dot denotes atom Ga, the blue-green dot denotes atom Al and the small gray dot atom N.

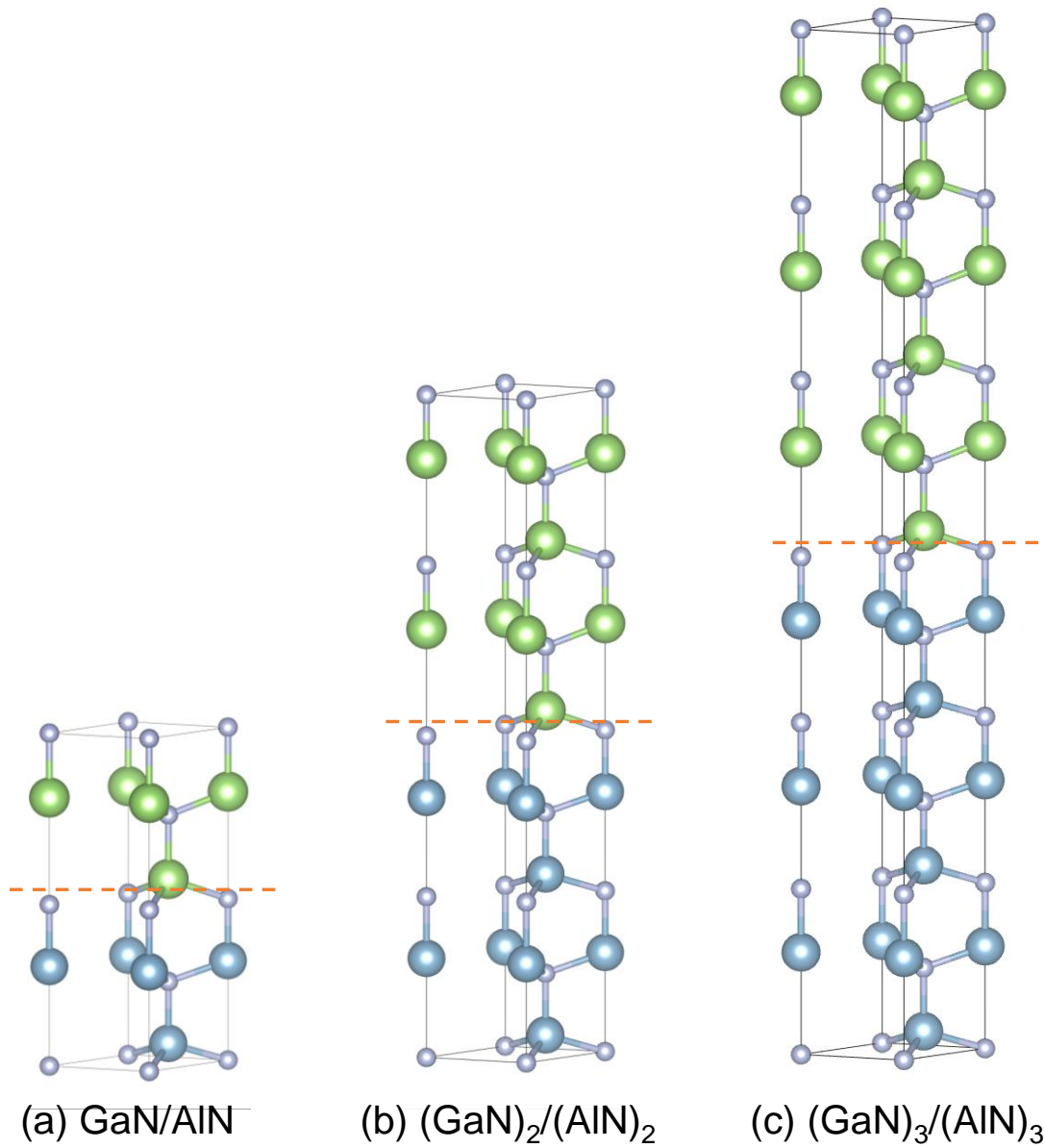

Figure 2 Lattice structures of superlattices $(GaN)_x/(AlN)_x$. The bright green dot denotes atom Ga, the blue-green dot denotes atom Al and the small gray dot atom N.

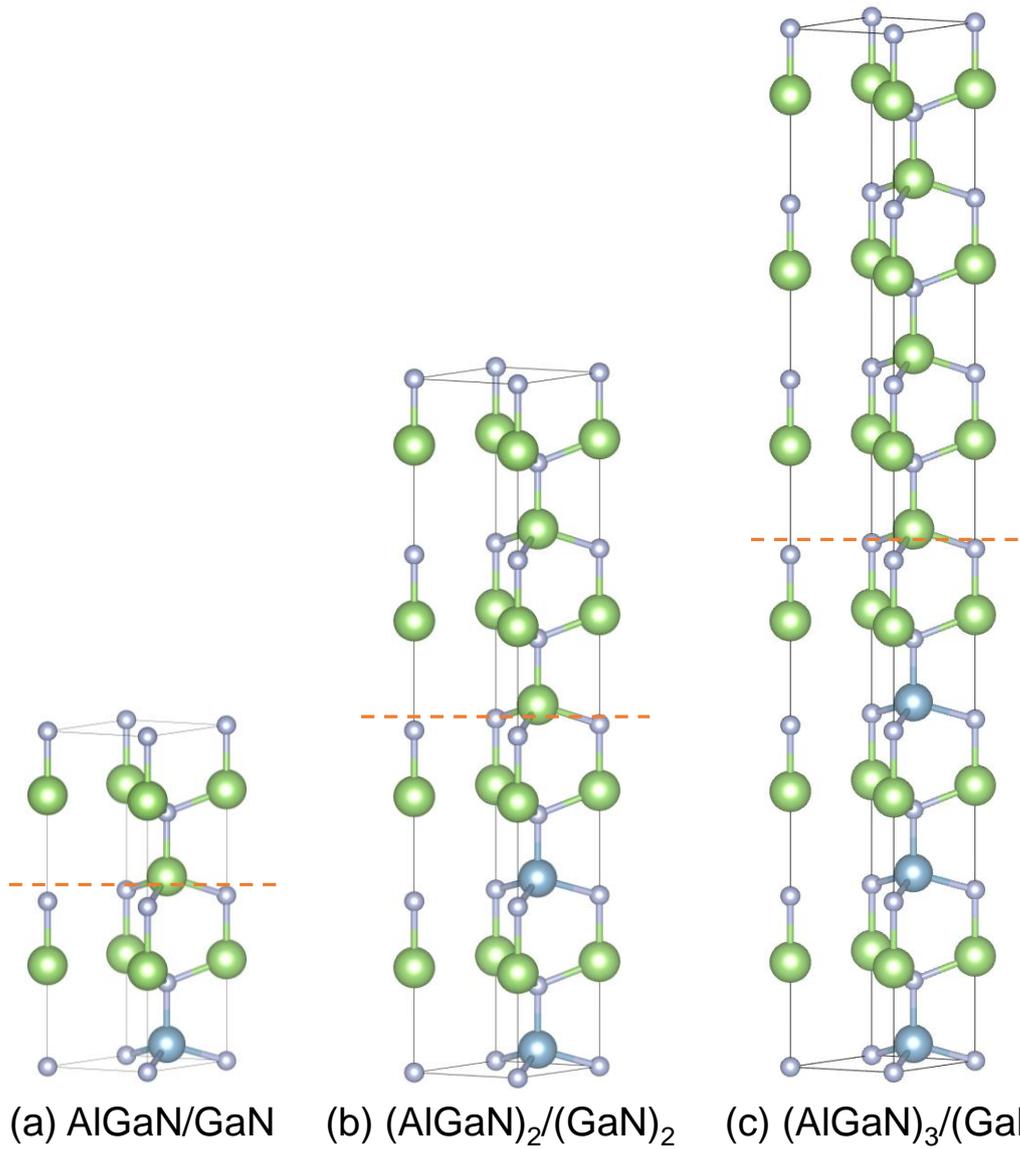

(a) AlGaN/GaN  (b) $(AlGaN)_2/(GaN)_2$  (c) $(AlGaN)_3/(GaN)_3$

Figure 3 Lattice structures of superlattices $(AlGaN)_x/(GaN)_x$. The bright green dot denotes atom Ga, the blue-green dot denotes atom Al and the small gray dot atom N.

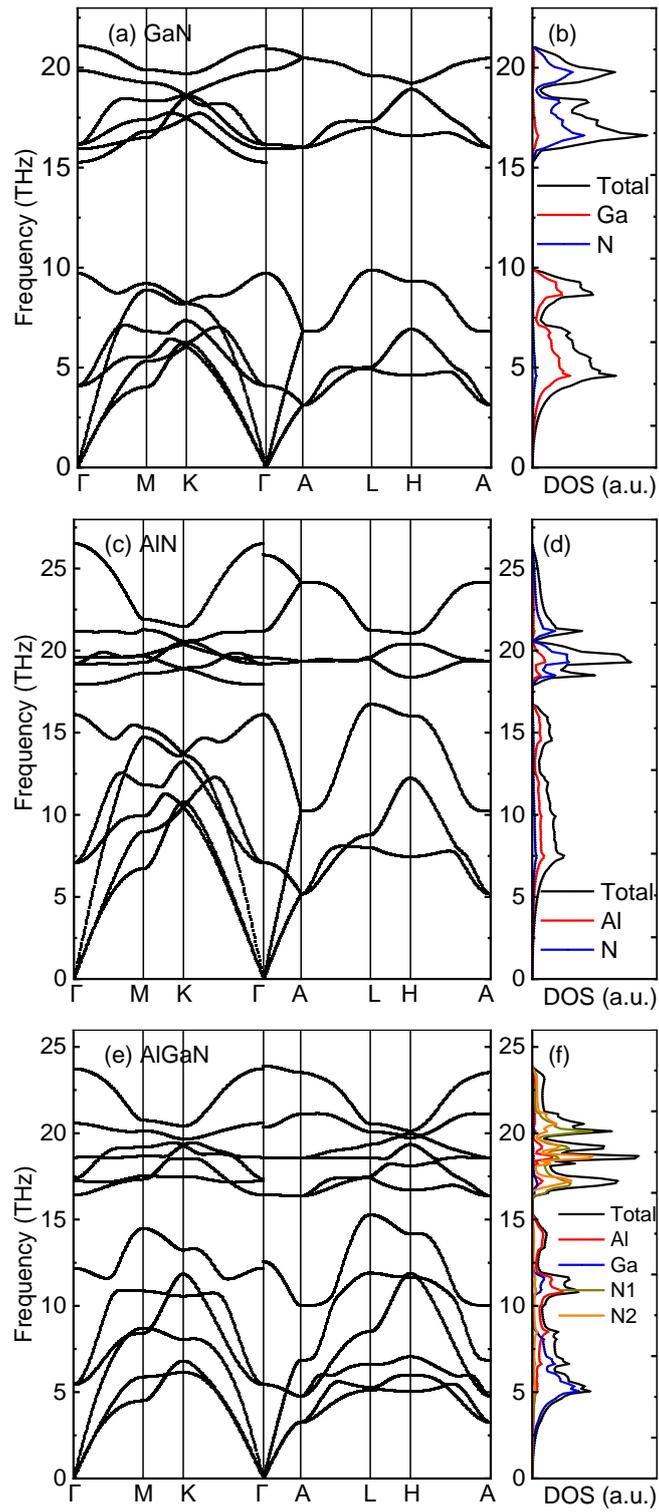

Figure 4 Phonon dispersions and density of states of (a)(b) GaN, (c)(d) AlN, (e)(f) AlGaN

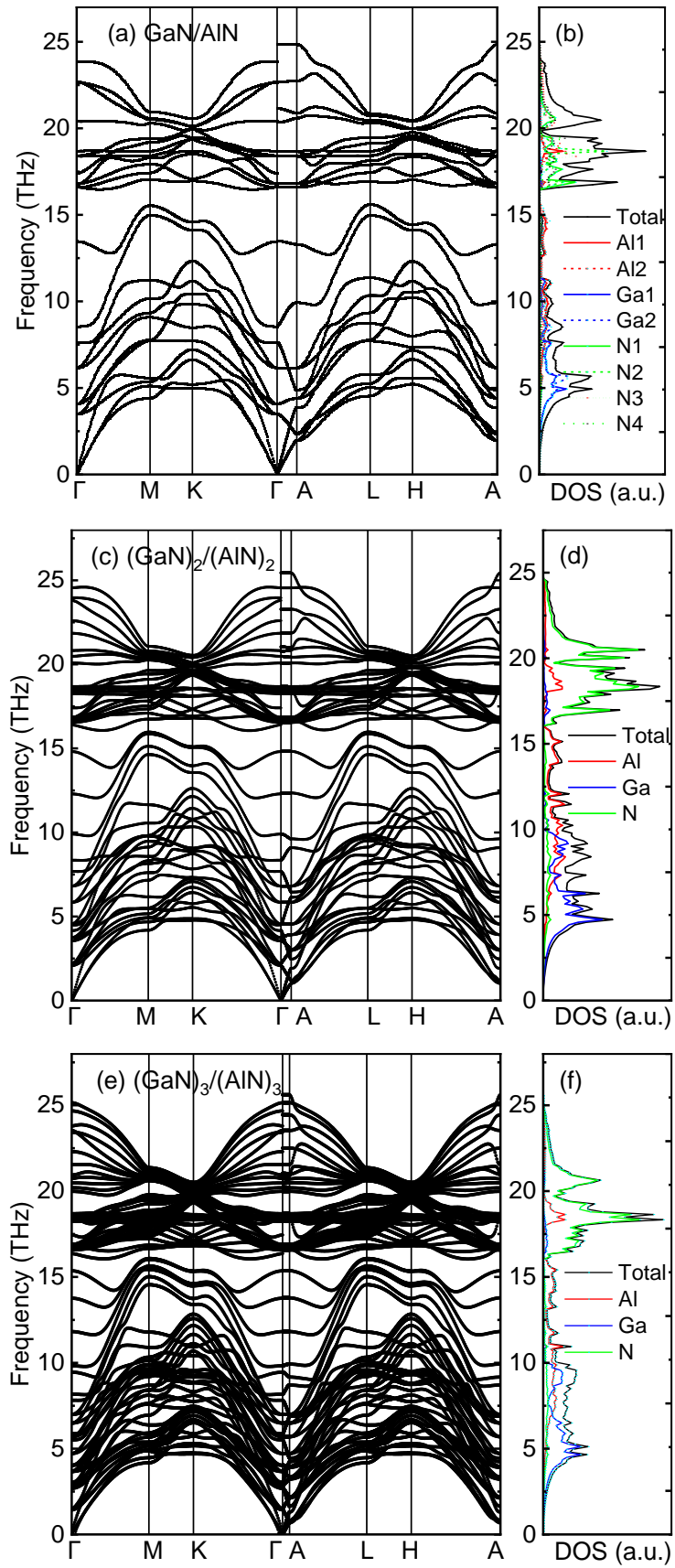

Figure 5 Phonon dispersions and density of states of (a)(b) GaN/AlN, (c)(d) $(GaN)_2/(AlN)_2$, and (e)(f) $(GaN)_3/(AlN)_3$

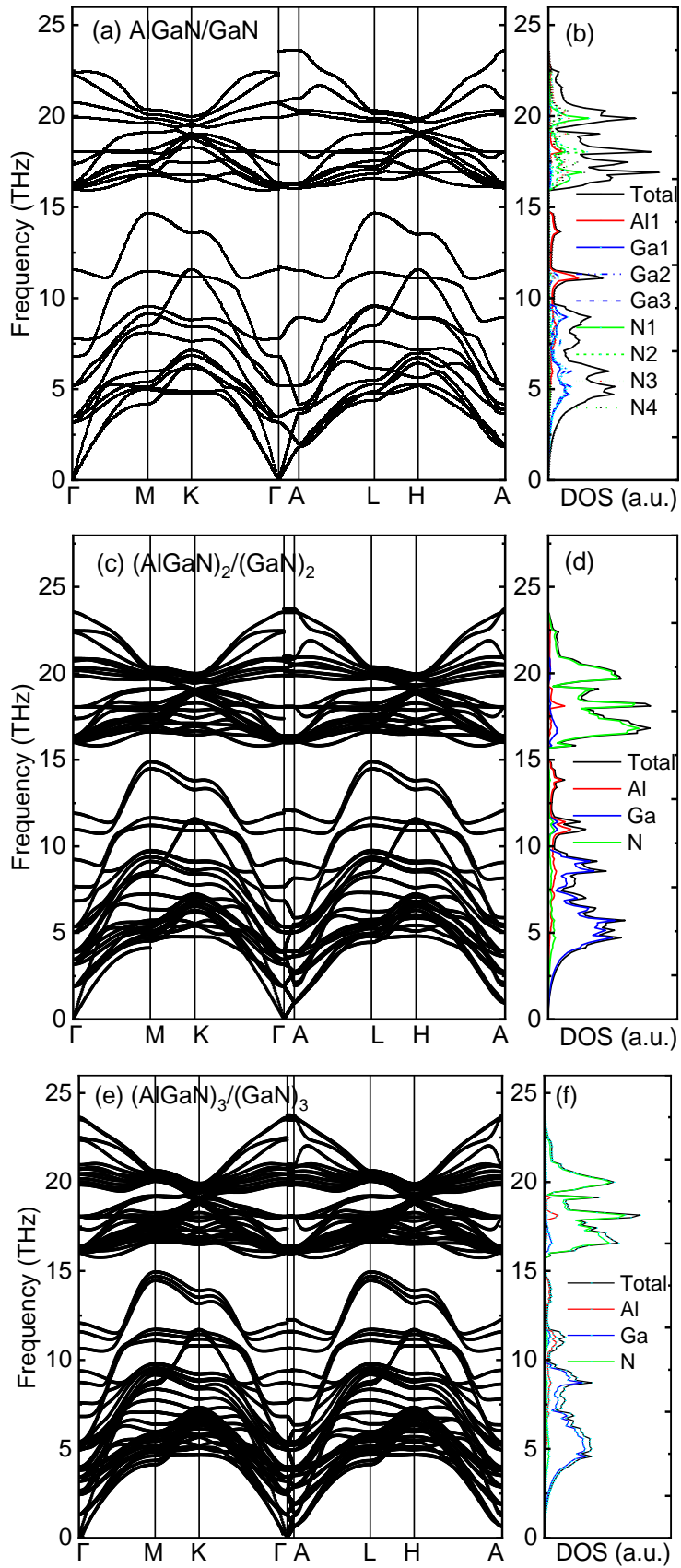

Figure 6 Phonon dispersions and density of states of (a)(b) AlGaN/GaN, (c)(d) (AlGaN)$_2$/(GaN)$_2$, and (e)(f) (AlGaN)$_3$/(GaN)$_3$

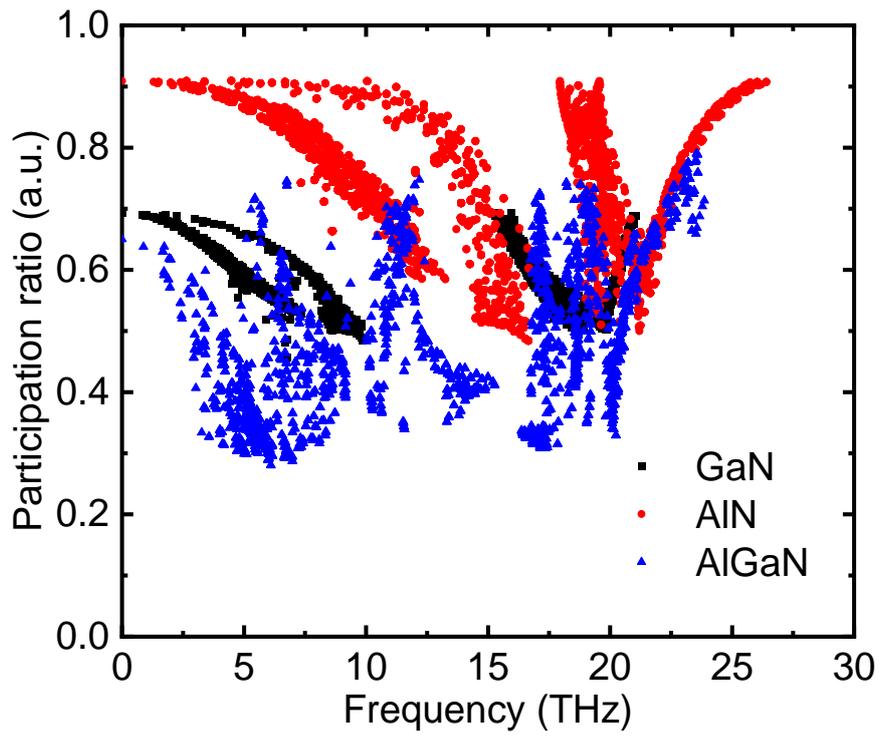

Figure 7 Phonon modal participation ratio in GaN, AlN, AlGaN

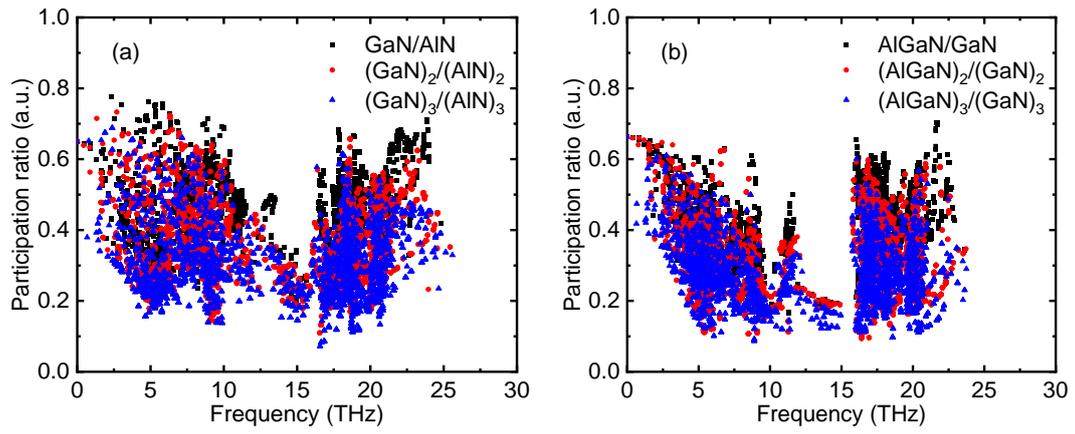

Figure 8 Phonon modal participation ratio in GaN superlattices: (a) $(GaN)_x/(AlN)_x$ and (b) $(AlGaN)_x/(GaN)_x$

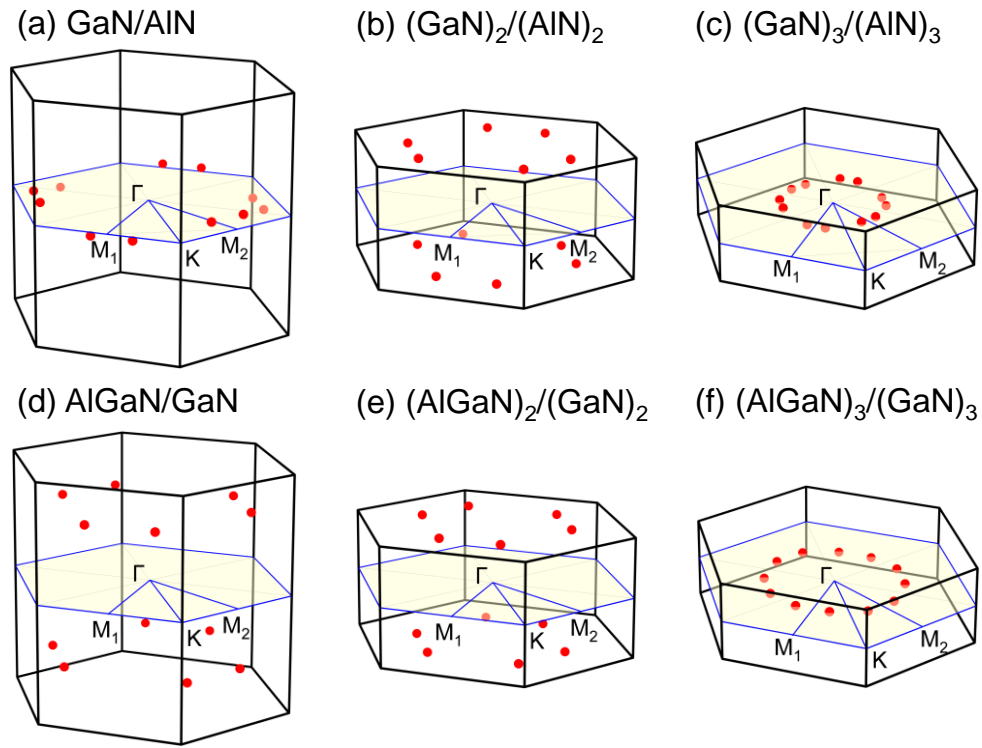

Figure 9 Brillouin zones of six GaN superlattice structures and the positions of Weyl phonons discussed in the main text. The black lines are used to indicate the Brillouin zone, blue lines for the $k_z=0$ plane, and high symmetry paths in this plane. The Weyl phonons are labeled with red dots.

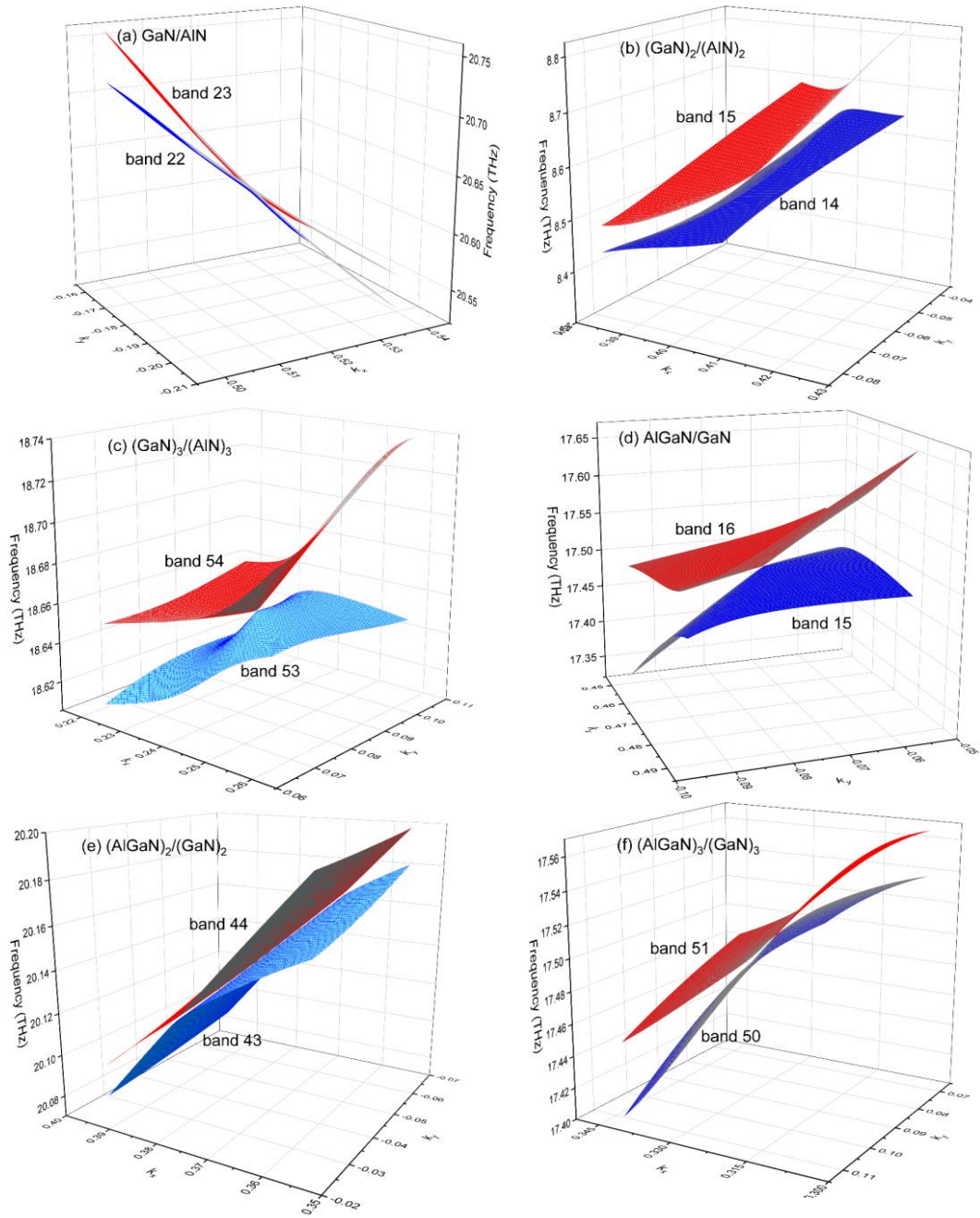

Figure 10 Three-dimensional phonon dispersions around the Weyl phonons in six superlattices marked in Figure 9.

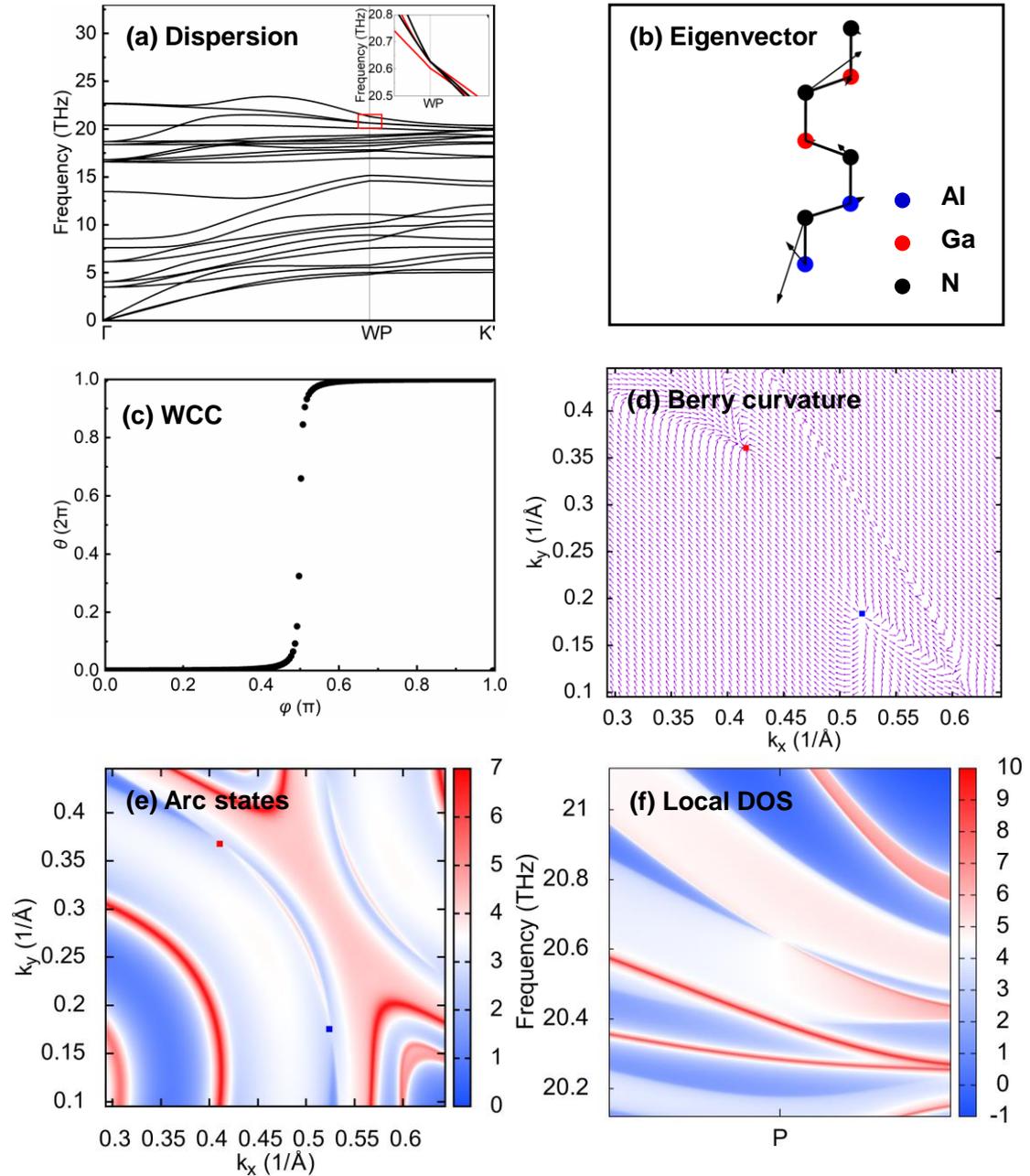

Figure 11 Weyl phonons at band 22 with **q** (0.399, 0.097, 0.032) at frequency 20.63 THz in GaN/AlN (a) Phonon dispersion (b) Phonon eigenvector (c) WCC evolution (d) Berry curvature (e) Surface arc state (f) Local destiny of states. The label K' in (a) denotes the **q** points K but with the same $k_z$ as the Weyl phonon, i.e., (1/3, 1/3, $k_0$). The black line and red line in the enlarged view of (a) are the phonon dispersions at the Weyl phonon point without and with NAC. The vibration vectors in (b) are enlarged for a better view.

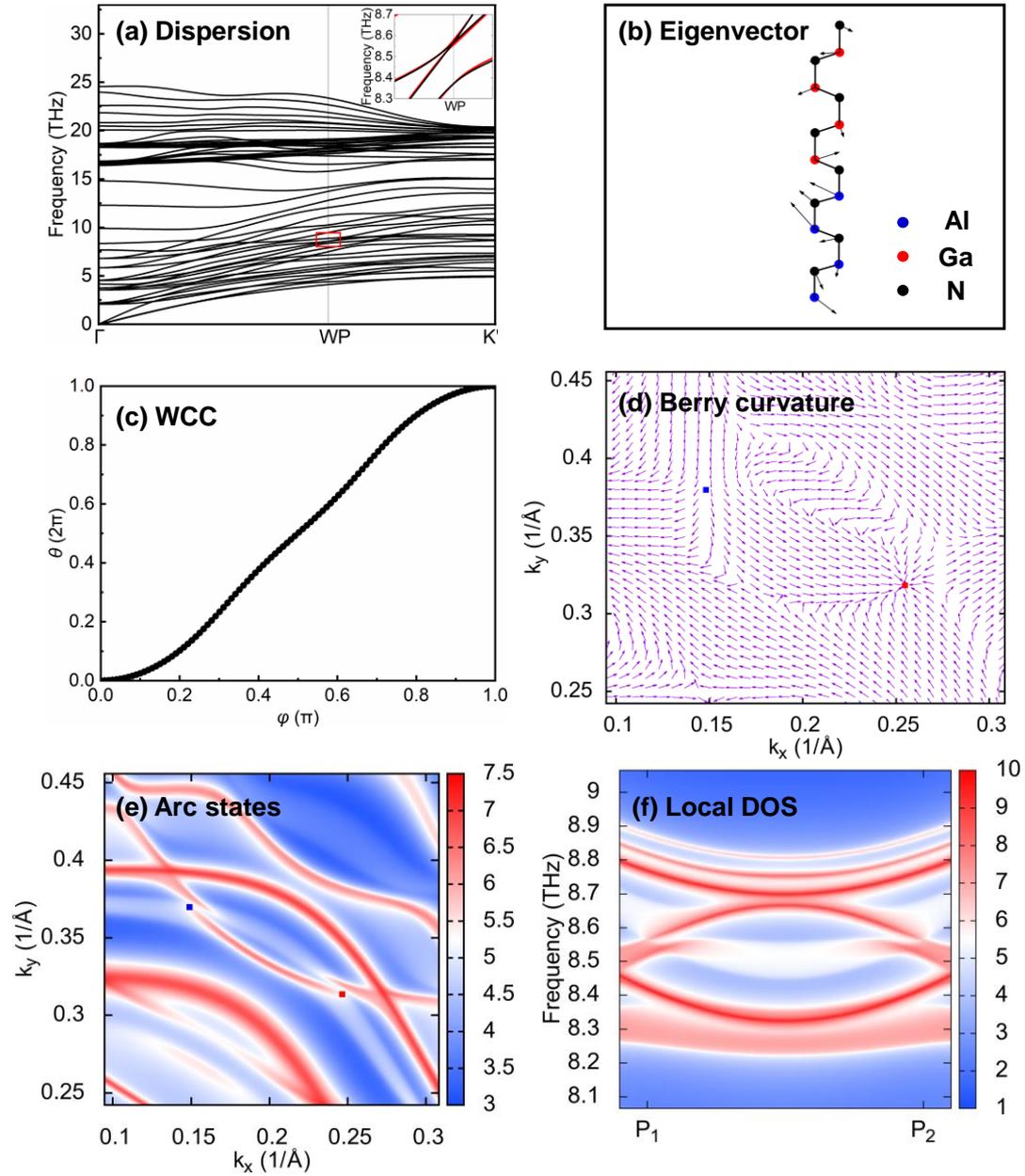

Figure 12 Weyl phonons at band 14 with **q** (0.243, 0.141, 0.463) at frequency 8.56 THz in $(GaN)_2/(AlN)_2$ (a) Phonon dispersion (b) Phonon eigenvector (c) WCC evolution (d) Berry curvature (e) Surface arc state (f) Local destiny of states. The label K' in (a) denotes the **q** points K but with the same $k_z$ as the Weyl phonon, *i.e.*, (1/3, 1/3, $k_0$). The black line and red line in the enlarged view of (a) are the phonon dispersions at the Weyl phonon point without and with NAC. The blue dot in (d) illustrates the position of the Weyl phonon at the $k_z = -k_0$ plane.

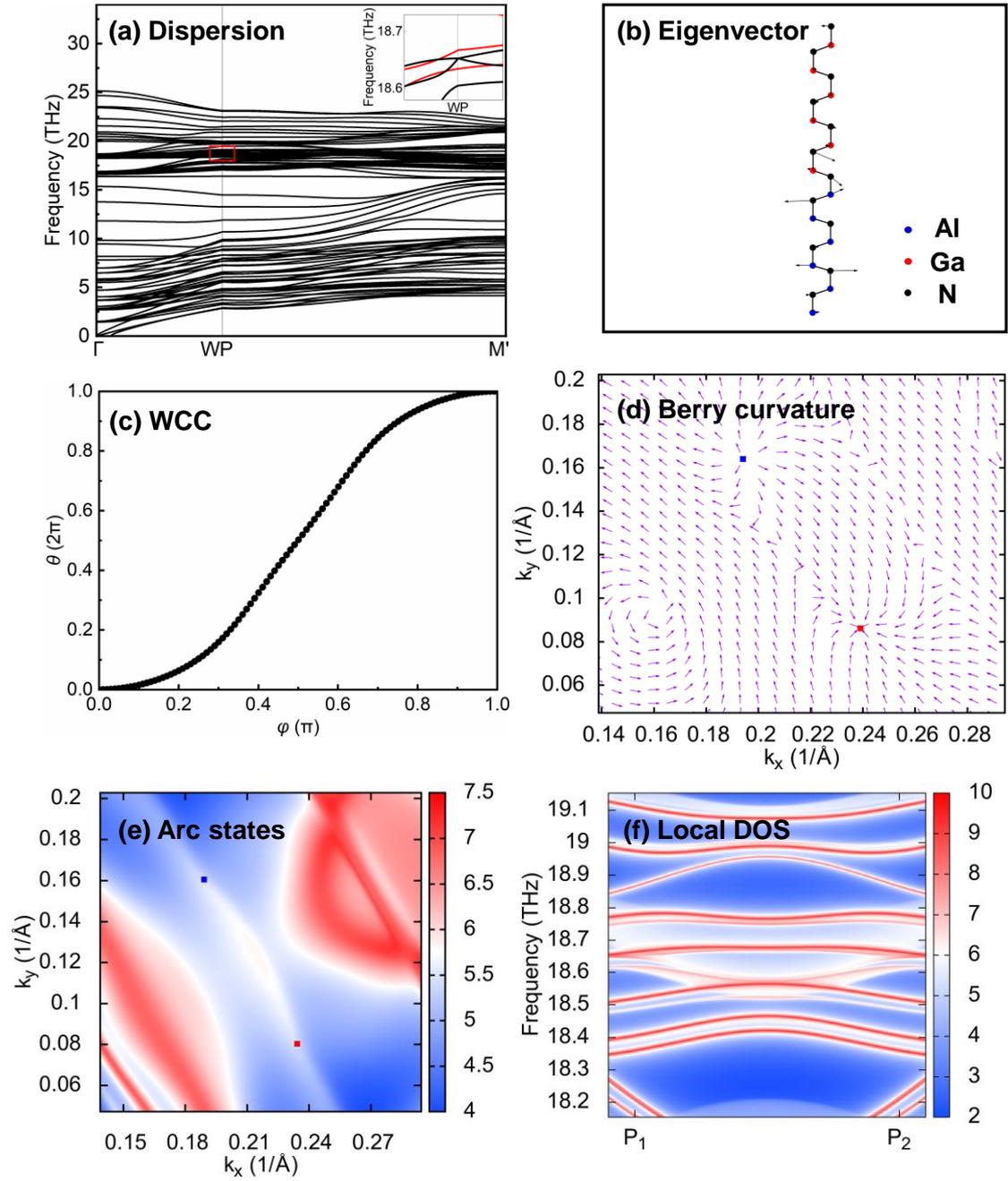

Figure 13 Weyl phonons at band 53 with **q** (0.043, 0.185, -0.027) at frequency 18.65 THz in $(GaN)_3/(AlN)_3$ (a) Phonon dispersion (b) Phonon eigenvector (c) WCC evolution (d) Berry curvature (e) Surface arc state (f) Local destiny of states. The label M' in (a) denotes the **q** points M but with the same $k_z$ as the Weyl phonon, *i.e.*, (1/3, 1/3, $k_0$). The black line and red line in the enlarged view of (a) are the phonon dispersions at the Weyl phonon point without and with NAC.

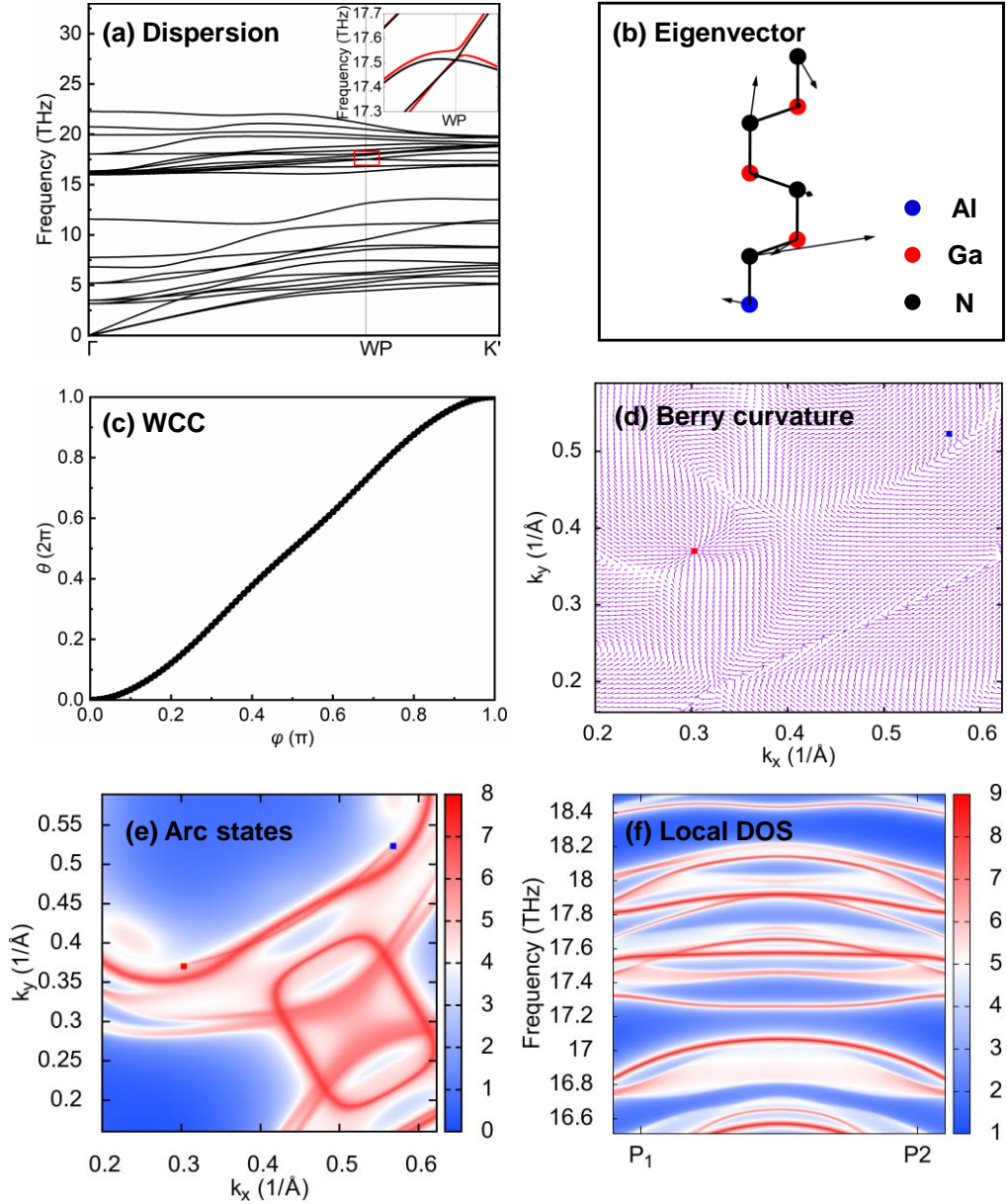

Figure 14 Weyl phonons at band 15 with **q** (0.291, 0.163, -0.313) at frequency 17.51 THz in AlGaN/GaN (a) Phonon dispersion (b) Phonon eigenvector (c) WCC evolution (d) Berry curvature (e) Surface arc state (f) Local destiny of states. The label K' in (a) denotes the **q** points K but with the same $k_z$ as the Weyl phonon, *i.e.*, (1/3, 1/3, $k_0$). The black line and red line in the enlarged view of (a) are the phonon dispersions at the Weyl phonon point without and with NAC. The blue dot in (d) illustrates the position of the Weyl phonon at the $k_z = -k_0$ plane.

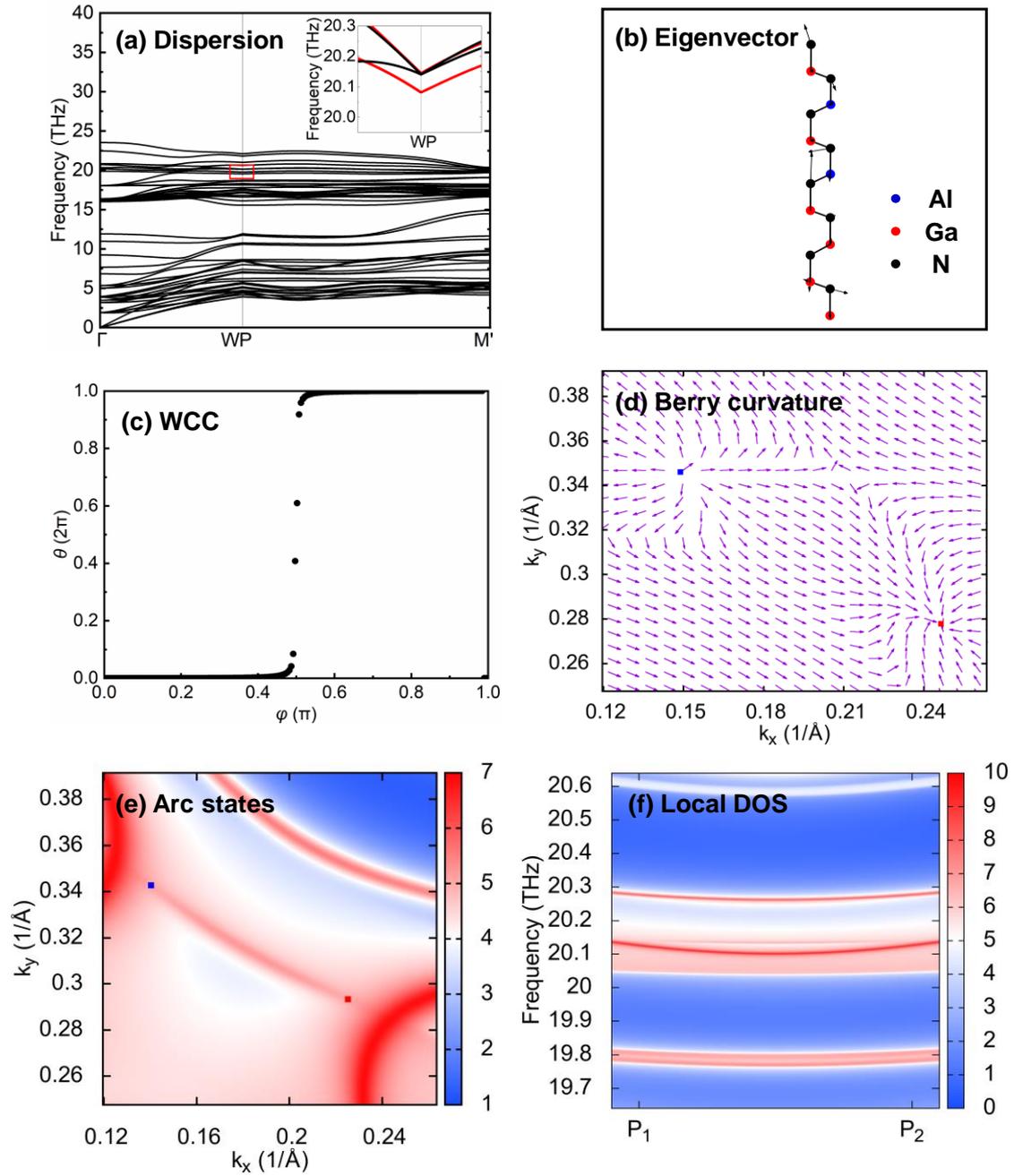

Figure 15 Weyl phonons at band 43 with **q** (0.217, 0.143, -0.495) at frequency 20.14 THz in $(AlGaN)_2/(GaN)_2$ (a) Phonon dispersion (b) Phonon eigenvector (c) WCC evolution (d) Berry curvature (e) Surface arc state (f) Local destiny of states. The label M' in (a) denotes the **q** points M but with the same $k_z$ as the Weyl phonon, *i.e.*, (1/3, 1/3, $k_0$). The black line and red line in the enlarged view of (a) are the phonon dispersions at the Weyl phonon point without and with NAC. The sink marked by the blue dot in (d) Berry curvature is the indication of the Weyl phonon at the $k=-k_0$ plane, which is also present in the Berry curvature at the $k=k_0$ plane due to the short distance between these two planes.

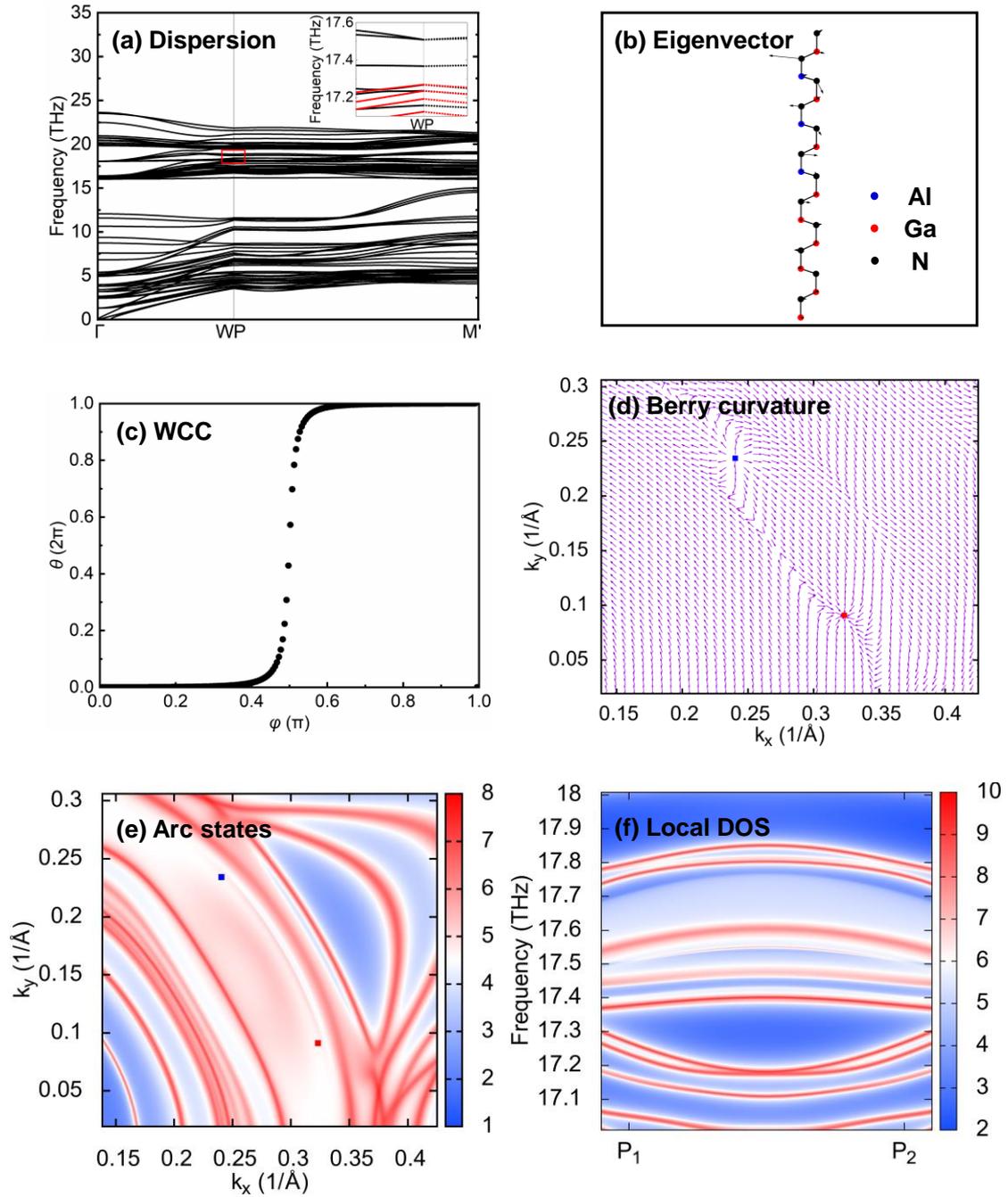

Figure 16 Weyl phonons at band 50 with **q** (0.080, 0.231, 0.011) at frequency 17.51 THz in $(AlGaN)_3/(GaN)_3$ (a) Phonon dispersion (b) Phonon eigenvector (c) WCC evolution (d) Berry curvature (e) Surface arc state (f) Local destiny of states. The label M' in (a) denotes the **q** points M but with the same $k_z$ as the Weyl phonon, *i.e.*, (1/3, 1/3, $k_0$). The black line and red line in the enlarged view of (a) are the phonon dispersions at the Weyl phonon point without and with NAC.

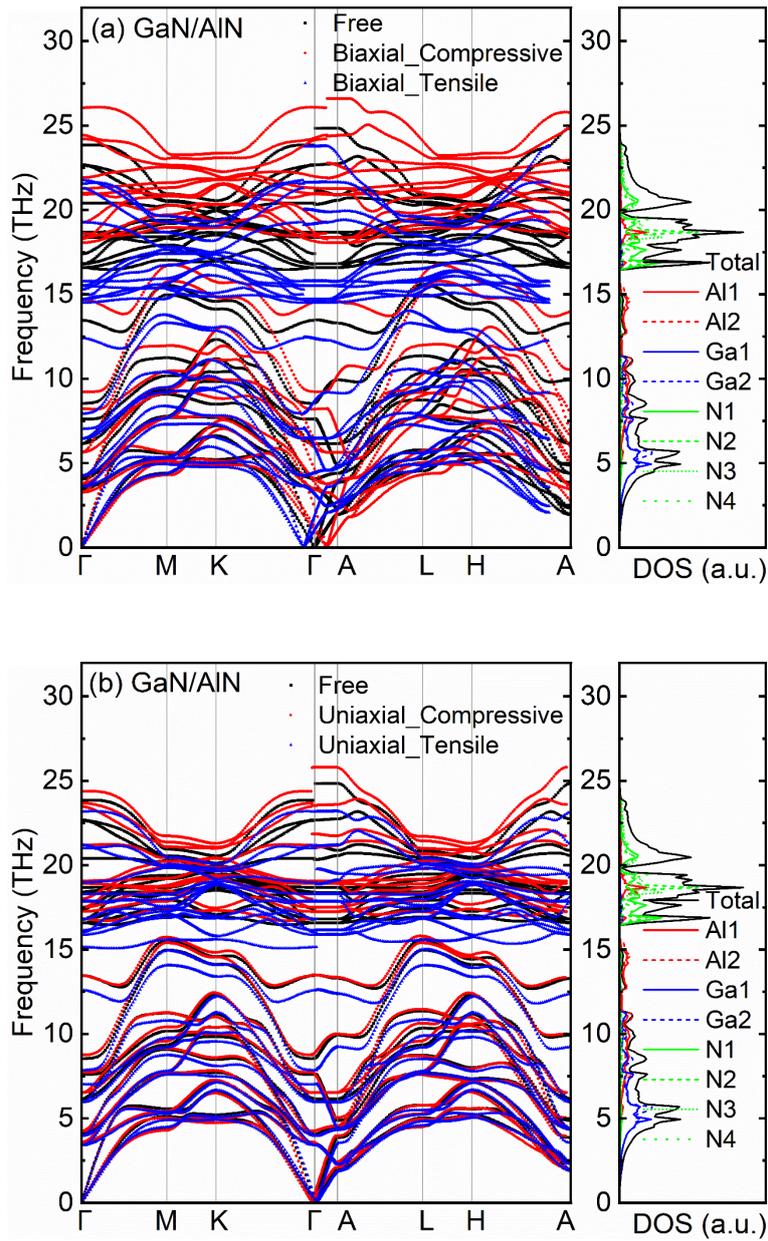

Figure 17 Phonon dispersions of GaN/AlN superlattice under different strain states (a) Compressive and tensile biaxial strain, (b) Compressive and tensile uniaxial strain along the polar axis.

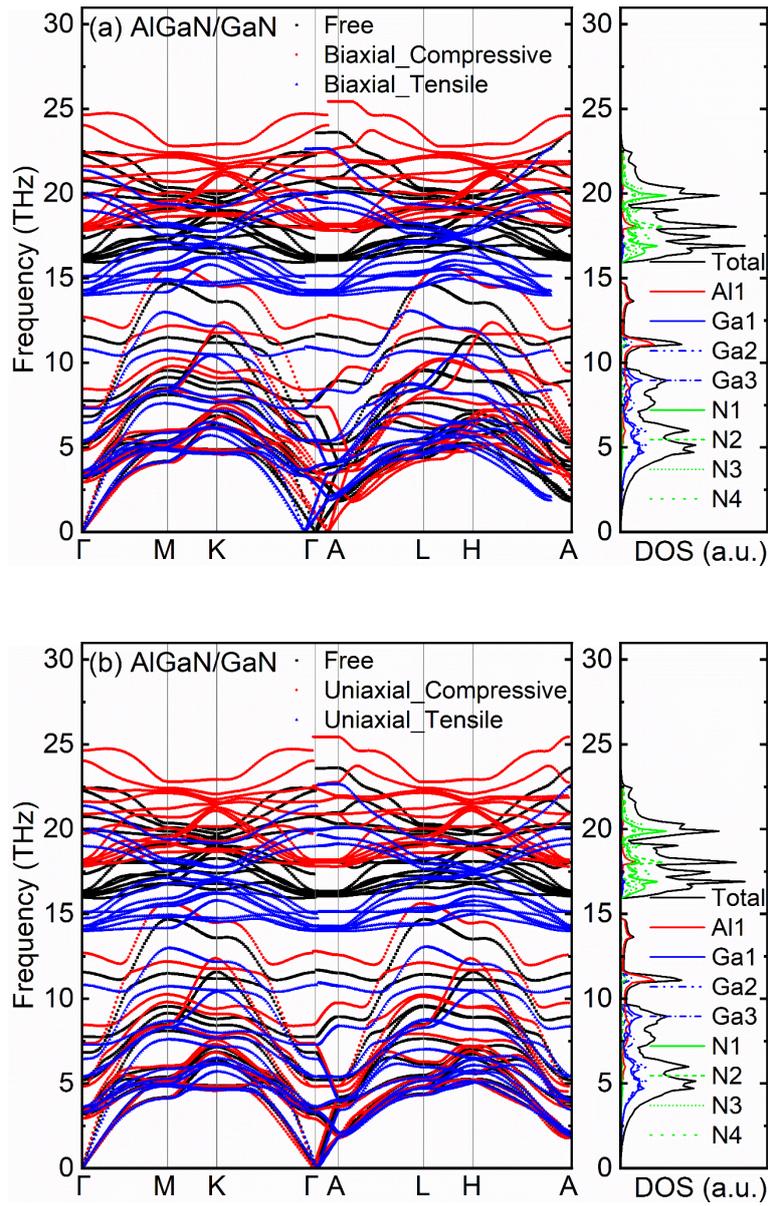

Figure 18 Phonon dispersions of AlGaN/GaN superlattice under different strain states (a) Compressive and tensile biaxial strain, (b) Compressive and tensile uniaxial strain along the polar axis.

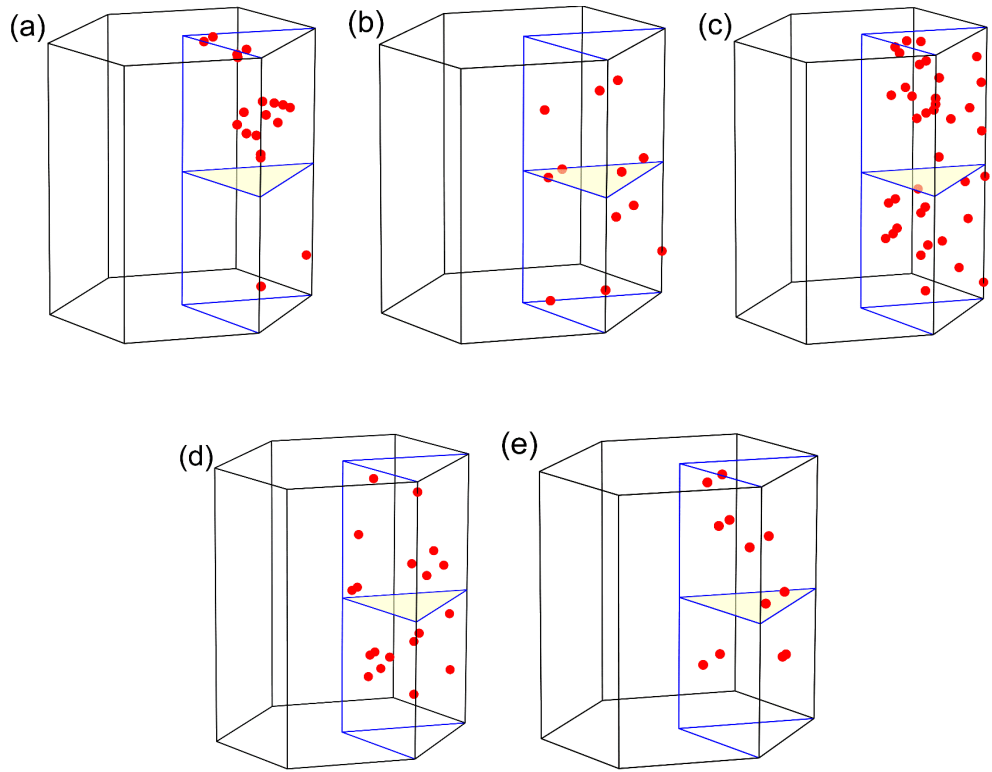

Figure 19. Weyl phonon points in irreducible Brillouin zone (marked by the blue lines) for GaN/AlN superlattice under different strain states (a) free, (b) biaxial compressive strain, (c) biaxial tensile stain, (d) uniaxial compressive strain, (e) uniaxial tensile strain.

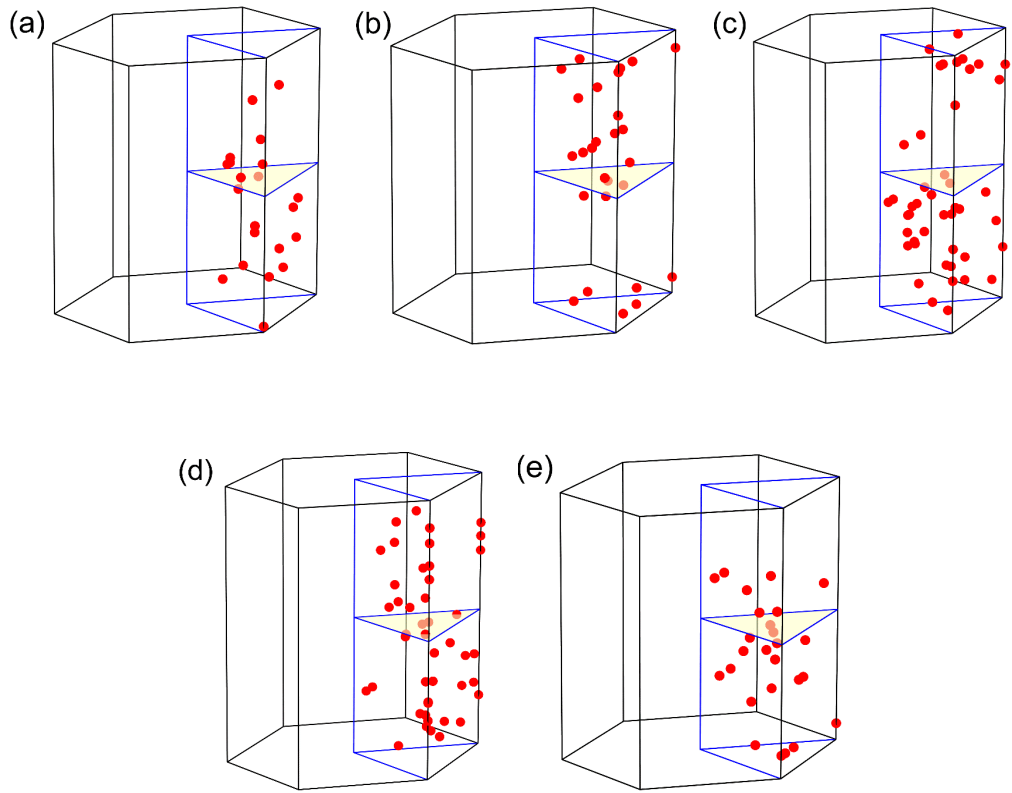

Figure 20. Weyl phonon points in irreducible Brillouin zone (marked by the blue lines) for AlGaN/GaN superlattice under different strain states (a) free, (b) biaxial compressive strain, (c) biaxial tensile stain, (d) uniaxial compressive strain, (e) uniaxial tensile strain.